\documentclass[aps,pre,reprint,twocolumn,superscriptaddress,showpacs]{revtex4-2}
\usepackage{amssymb,amsmath}
\usepackage[table]{xcolor}
\usepackage{graphicx}
\usepackage{mathtools}
\usepackage[export]{adjustbox}
\usepackage{overpic}
\usepackage[colorlinks=true,
linkcolor=black,
urlcolor=blue,
citecolor=blue]{hyperref}
\usepackage{nicefrac}
\usepackage{multirow}
\usepackage{lipsum} 
\usepackage[normalem]{ulem}
\usepackage{pbox}
\newcolumntype{C}[1]{>{\centering\let\newline\\\arraybackslash\hspace{0pt}}m{#1}}
\usepackage{verbatim}
\usepackage{enumitem}

\usepackage{framed,color}
\definecolor{shadecolor}{rgb}{0.85,0.80,0.80}

\definecolor{myorange}{RGB}{253, 184, 99}
\definecolor{mypurple}{RGB}{178, 171, 210}
\newcommand\numberthis{\addtocounter{equation}{1}\tag{\theequation}}

\newcommand{\comments}[1]{}
\usepackage{multirow}

\usepackage{float}

\newcommand{\beq}{\begin{equation}}
	\newcommand{\eeq}{\end{equation}}
\newcommand{\bal}{\begin{aligned}}
	\newcommand{\eal}{\end{aligned}}

\newcommand{\be}{\begin{equation}}
	\newcommand{\ee}{\end{equation}}
\newcommand{\bd}{\begin{displaymath}}
	\newcommand{\ed}{\end{displaymath}}
\newcommand{\BE}{\begin{eqnarray}}
	\newcommand{\EE}{\end{eqnarray}}

\newcommand{\id}{{\openone}}

\newcommand{\jb}[1]{\textbf{\textcolor{green}{[JB: #1]}}}

\allowdisplaybreaks
\begin{document}
	\title{Eigenvalue spectra and stability of directed complex networks}
	\author{Joseph W. Baron}
	\email{josephbaron@ifisc.uib-scic.es}
	\affiliation{Instituto de F{\' i}sica Interdisciplinar y Sistemas Complejos IFISC (CSIC-UIB), 07122 Palma de Mallorca, Spain}

	\begin{abstract}
		Quantifying the eigenvalue spectra of large random matrices allows one to understand the factors that contribute to the stability of dynamical systems with many interacting components. This work explores the effect that the interaction network between components has on the eigenvalue spectrum. We build upon previous results, which usually only take into account the mean degree of the network, by allowing for non-trivial network degree heterogeneity. We derive closed-form expressions for the eigenvalue spectrum of the adjacency matrix of a general weighted and directed network. Using these results, which are valid for any large well-connected complex network, we then derive compact formulae for the corrections (due to non-zero network heterogeneity) to well-known results in random matrix theory. Specifically, we derive modified versions of the Wigner semi-circle law, the Girko circle law and the elliptic law and any outlier eigenvalues. We also derive a surprisingly neat analytical expression for the eigenvalue density of an directed Barabasi-Albert network. We are thus able to make general deductions about the effect of network heterogeneity on the stability of complex dynamic systems.
	\end{abstract}
	
	
	
	\maketitle
	
	\section{Introduction}
	Random matrix theory (RMT) is an area of study in complex systems with a diverse range of applications \cite{RMPhysicsBook, taobook, akemann2011oxford}. This largely owes to the generality of the question that it attempts to answer. Namely, if one linearises a dynamical system with many components about a fixed point, what are the statistics of the entries of the Jacobian matrix that permit stability? Consequently, RMT is an important theoretical tool in subjects as wide-ranging as neural networks \cite{aljadeff2015transition, kuczala2016eigenvalue, rajan2006eigenvalue, coolen2005theory}, spin-glasses \cite{mezard1987, braymoore, kosterlitz1976spherical}, finance \cite{laloux2000random}, telecommunications theory \cite{tulino2004random} and theoretical ecology \cite{may, allesinatang1, allesinatang2, gravel, barabas2017, allesina2015}.

	Motivated by these applications, the selection of random matrix ensembles being investigated is rapidly broadening. In recent years, random matrices with block-structure \cite{baron2020dispersal, allesina2015predicting, gravel}, cyclic correlations \cite{rogers2010}, generalised correlations \cite{baron2022eigenvalues}, element-specific variances and correlations \cite{aljadeff2015transition, kuczala2016eigenvalue}, sum and product structures \cite{rogers2010universal, ahmadian2015properties, gibbs2018effect}, and heterogeneous diagonal elements \cite{barabas2017}, amongst many other things, have all been examined. In all the above cases however, either all matrix elements were assumed to be non-zero (corresponding to fully-connected graph), or pairs of elements $(a_{ij},a_{ji})$ were assumed to be non-zero with a probability $C$, which does not scale with the matrix size $N$ ( corresponding to an Erd\"os-R\'enyi graph \cite{erdos1960evolution} with mean degree $CN$).

	However, progress has also been made in studying random matrices that correspond to more intricate complex network \cite{albert2002statistical} structures. For example, in-roads have been made towards understanding the spectra of sparse random networks, where the mean degree of the network is $p$, where $p$ does not scale with $N$. In this context, most attention has been paid to sparse Erd\"os-R\'enyi graphs \cite{semerjian2002sparse, rodgers1988density, biroli1999single}, but some works have also endeavoured to quantify the spectra of scale-free networks \cite{anand2011shannon, rodgers2005eigenvalue, nagao2008spectral} or the Cayley tree \cite{kim1985density, bordenave2010resolvent}, for example, and also more generally \cite{metz2020spectral}. In the aforementioned works, the matrices in question were symmetric, but efforts have also been made to find the spectra of asymmetric sparse random matrices (again with Erd\"os-R\'enyi or tree-like structure) \cite{rogers2009cavity, neri2012spectra}. Most notably, some recent success has been had in finding the spectra of asymmetric random matrices with more intricate network structure \cite{metz2019spectral, tarnowski2020universal, metz2021localization}, but this often requires the use of computationally expensive algorithms and has largely not yielded analytical or closed-form results. 
	
	In this work, we analytically deduce the eigenvalue spectra of asymmetric random matrices with general dense complex network structure. We find closed-form expressions for the eigenvalue density, the support of the eigenvalue spectrum and the location of any outlier eigenvalues. We then use these expressions to deduce succinct approximations in the case of small network heterogeneity. This allows us to show explicitly how the well-known elliptic \cite{sommers}, Girko circle \cite{girko1985circular} and Wigner semi-circle \cite{wigner1958distribution, wigner1967random} laws are modified by the introduction of a non-trivial complex network structure (i.e. differing from the Erd\"os-R\'enyi graph). Further, in the case where transpose pairs of elements are uncorrelated, we derive an exact universal modified circular law, which is valid for arbitrarily large network heterogeneity. Finally, we also show how our results can be used to provide exact expressions for the eigenvalue density of some complex networks. In particular, we provide a simple analytical expression for the eigenvalue density of a Barabasi-Albert network \cite{albert2002statistical}.
	
	The rest of this work is structured as follows. In Section \ref{section:randommatrixensemble}, we introduce the random matrix ensemble with which we will be working. In Section \ref{section:annealedcalculation}, we describe the method that is used to deduce the eigenvalue density of this random matrix ensemble. In Section \ref{section:generalresults}, we present results for the eigenvalue density, support of the eigenvalue spectrum and the locations of any outliers. These results are valid for any dense, large random matrix with general complex network structure. Then, in Section \ref{section:corrections}, we show how these general formulae can be approximated for small network heterogeneity (or evaluated exactly in some cases) and derive the aforementioned modified circular, elliptical and semi-circular laws. In Section \ref{section:banetwork}, we derive a simple expression for the eigenvalue density of a directed Barabsi-Albert network. Given all these results, in Section \ref{section:stability}, we summarise and interpret our results and discuss the effect that network degree heterogeneity has on the stability of complex systems in general. Finally, we discuss possibilities for extending and applying the methods developed here in Section \ref{section:conclusion}.

	\section{Random matrix ensemble}\label{section:randommatrixensemble}
	\subsection{Directed complex network as an asymmetric random matrix}
	A weighted and directed complex network is represented by a random matrix with elements $a_{ij}$. A non-zero value of $a_{ij}$ indicates an edge connecting nodes $i$ and $j$. The value of $a_{ij}$ indicates the weight of that edge, which may be negative. If $a_{ij}$ is non-zero, then $a_{ji}$ is also non-zero, but in general $a_{ij} \neq a_{ji}$. In the network, there are a total number of nodes $N$, and the nodes have an average degree $p$. 
	
	We imagine that the matrix $\underline{\underline{a}}$ is constructed by drawing diagonally opposite pairs of elements $(a_{ij}, a_{ji})$ from the following joint distribution
	\begin{align}
		P(a_{ij}, a_{ji}) = (1 - f_{ij}) \delta(a_{ij})\delta(a_{ji}) + f_{ij} \pi(a_{ij}, a_{ji}), \label{matrixensemble}
	\end{align}
	where $f_{ij}$ is the probability that the elements $a_{ij}$ and $a_{ji}$ are non-zero and where the joint distribution of non-zero entries $\pi(a_{ij}, a_{ji})$ has statistics given by
	\begin{align}
		\left\langle a_{ij} \right\rangle_\pi &= \frac{\mu}{p} , \nonumber \\
		\left\langle \left(a_{ij} - \frac{\mu}{p} \right)^2 \right\rangle_\pi &= \frac{\sigma^2}{p} , \nonumber \\
		\left\langle \left(a_{ij} - \frac{\mu}{p} \right) \left(a_{ji} - \frac{\mu}{p} \right)\right\rangle_\pi &= \frac{\Gamma\sigma^2}{p} , \label{pimoments}
	\end{align}
	where the notation $\langle \cdot \rangle_{\pi}$ indicates an average over the distribution $\pi(a_{ij}, a_{ji})$. This scaling of the moments with $p$ ensures a sensible limit $p\to \infty$ \cite{mezard1987}. We presume that all higher-order moments and correlations decay more quickly than $p^{-1}$, which means they can be ignored [see Supplemental Material (SM) Section S2]. We also tacitly presume, in formulating Eq.~(\ref{matrixensemble}) in this way, that the degree of a node and the weights of the edges that connect to it are independent. 
	
	\subsection{Probability of connection}
	The coefficients $f_{ij}$ in Eq.~(\ref{matrixensemble}), the probabilities that nodes $i$ and $j$ are connected, determine the structure of the network. To approximate $f_{ij}$, we ignore correlations between node degrees and we assume that the network and its properties are determined by the degree distribution $\gamma(k)$, the probability that a randomly selected node has degree $k$.
	
	Suppose we are given the degree sequence $\{k_i\}$ of the network. There are a total number $pN/2$ of links in the network. In a similar fashion to the configuration model \cite{newman2018networks}, we then imagine that we pair all the nodes together, choosing nodes with a probability proportional to their degree. The probability that nodes $i$ and $j$ are paired is therefore \cite{rodgers2005eigenvalue}
	\begin{align}
		f_{ij} &= 1 - \left[1 - 2 \frac{k_i k_j}{(pN)^2}\right]^{pN/2} \approx 1 - \exp\left(- \frac{k_i k_j}{pN}\right) \nonumber \\
		&\approx \frac{k_i k_j}{pN} , \label{fapprox}
	\end{align}
	where we assume $\frac{k_i k_j}{pN} \ll 1$. We emphasize at this point that the network in question need not necessarily have been constructed according to the configuration model. The final results for the eigenvalue spectrum, which are accurate for large $N$ and $p$, only require information about the degree distribution $\gamma(k)$ and not the degrees of individual nodes. This is discussed further in Section \ref{section:annealedandblock}.
	
	\section{Calculating the eigenvalue spectrum}\label{section:annealedcalculation}
	In this section, we give a brief summary of the techniques we use to calculate the eigenvalue spectrum of the matrix $\underline{\underline{a}}$. A more thorough account of this calculation is given in SM Section S2.
	\subsection{Eigenvalue potential and resolvent}
	The eigenvalue spectrum, examples of which can be seen in Figs. \ref{fig:modifiedellipse} and \ref{fig:universality}, has two contributions: (1) a bulk region to which most of the eigenvalues are confined, (2) a single outlier that arises a non-zero value of $\mu$. That is,
	\begin{align}
		\rho(\omega) &= \left\langle\frac{1}{N} \sum_i \delta(\omega - \lambda_i) \right\rangle \nonumber \\
		&= \rho_{\mathrm{bulk}}(\omega) + \frac{1}{N} \delta(\omega -\lambda_{\mathrm{outlier}}), \label{densitydef}
	\end{align}
	where here angular brackets without a subscript $\pi$ indicate an average over the distribution $P(\{a_{ij}\}) = \prod_{i<j} P(a_{ij}, a_{ji})$ [see Eq.~(\ref{matrixensemble})]. We note that Eq.~(\ref{densitydef}) is normalised such that $\int d^2 \omega \rho(\omega) = 1$, where the integral is taken over the entire area of the complex plane. 
	
	Both the density of the eigenvalues in the bulk region and the location of the outlier eigenvalue can be calculated from the disorder-averaged resolvent \cite{sommers, benaych2011eigenvalues}
	\begin{align}
		\underline{\underline{G}}(\omega, \omega^\star) = \left\langle \left[\omega \underline{\underline{\id}} - \underline{\underline{a}} \right]^{-1}\right\rangle, \label{resolventdef}
	\end{align}
	where we note that the resolvent is not an analytic function of $\omega$ in areas of the complex plane where $\underline{\underline{z}}$ has a non-zero eigenvalue density \cite{haake, sommers}. Letting $G = N^{-1}\mathrm{Tr}\underline{\underline{G}}$, the eigenvalue density in the bulk region is given by 
	\begin{align}
		\rho_{\mathrm{bulk}}(\omega) = \frac{1}{\pi } \mathrm{Re} \left[\frac{\partial G}{\partial \omega^\star} \right].\label{bulkdensity}
	\end{align}
	The outlier eigenvalue can also be calculated in a similarly straightforward way from the resolvent matrix (see Section \ref{section:outliers}). 
	
	The resolvent matrix can be derived from the so-called eigenvalue `potential' \cite{sommers}
	\begin{align}
		\Phi(\omega, \omega^\star) = - \frac{1}{N} \left\langle\ln \det\left[(\omega^\star \underline{\underline{\id}} - \underline{\underline{a}}^T)(\omega \underline{\underline{\id}} - \underline{\underline{a}}) \right]\right\rangle,
	\end{align}
	which gives the trace of the resolvent via
	\begin{align}
		G = \frac{\partial \Phi}{\partial \omega} .
	\end{align}
	All the information we require about the eigenvalue spectrum is contained in the eigenvalue potential $\Phi(\omega, \omega^\star)$. That is, by evaluating $\Phi(\omega, \omega^\star)$, we are able to find the eigenvalue spectrum of $\underline{\underline{a}}$. In practice, $\Phi(\omega, \omega^\star)$ is evaluated using a saddle-point approximation, assuming large $N$ (see SM Section S2 and S3 for the full calculation).
	
	\subsection{Annealed network and correspondence to block-structured problem}\label{section:annealedandblock}
	We show in SM Section S2 that, for large $p$, the eigenvalue potential $\Phi(\omega, \omega^\star)$ can be approximated by that of an appropriately-weighted fully-connected graph.
	
	More precisely, we calculate $\Phi(\omega, \omega^\star)$ using the statistics for $a_{ij}$ given in Eq.~(\ref{matrixensemble}). For large $p$ and $p/N \ll 1$, we show that the same expression for $\Phi(\omega, \omega^\star)$ is obtained by drawing the pairs of elements from a joint distribution with the following statistics
	\begin{align}
		\left\langle a_{ij} \right\rangle &= \frac{k_i k_j}{pN}\frac{\mu}{p} , \nonumber \\
		\left\langle \left(a_{ij} - \frac{\mu}{N} \right)^2 \right\rangle &= \frac{k_i k_j}{pN}\frac{\sigma^2}{p} , \nonumber \\
		\left\langle \left(a_{ij} - \frac{\mu}{N} \right) \left(a_{ji} - \frac{\mu}{N} \right)\right\rangle &= \frac{k_i k_j}{pN}\frac{\Gamma\sigma^2}{p} . \label{annealedstatistics}
	\end{align}
	We note that here the angular brackets indicate an average that is not conditioned on the elements $(a_{ij}, a_{ji})$ being non-zero, unlike in Eq.~(\ref{pimoments}). Thus, the effect of the network on the eigenvalue spectrum is equivalent to that of an appropriately-weighted fully-connected graph. This is known as the annealed network approximation \cite{carro2016noisy}. 
	
	We now imagine that we organise the random matrix $\underline{\underline{a}}$ such that the rows and columns corresponding to nodes with the same degree $k$ are adjacent in the matrix (this does not change the eigenvalues of the matrix). Given the approximation for the statistics of the random matrix $\underline{\underline{a}}$ in  Eqs.~(\ref{annealedstatistics}), the problem reduces to that of calculating the eigenvalue spectrum of a random matrix with block-structured means and variances. 
	
	Random matrix problems with block structure have previously been addressed in the literature (see Refs. \cite{baron2020dispersal, kuczala2016eigenvalue, aljadeff2015transition} in particular). The reduction of the original complex-network ensemble to a block-structured random matrix ensemble thus enables us to deduce far more manageable analytical expressions for the eigenvalue spectrum. 
	
	\section{General results}\label{section:generalresults}
	In this Section, we provide general results for the bulk region of the eigenvalue spectrum and the outlier eigenvalues that would be valid for any uncorrelated network with $p/N \ll 1$ and sufficiently large $p$. Later in Section \ref{section:corrections}, we show how these more general results can be approximated (and therefore simplified drastically) in the case where the network degree heterogeneity is small. In Section \ref{section:banetwork}, we show also how the following general results simplify in the case of a Barabasi-Albert network (no small-heterogeneity approximation is made in this case).
	\subsection{Bulk region of the eigenvalue spectrum}\label{section:bulk}
	
	\subsubsection{Boundary of the bulk region}
	As is exemplified in Fig. \ref{fig:modifiedellipse}, the majority of the eigenvalues are confined to a bulk region of the complex plane. In SM Section S4 B, we demonstrate that the boundary of this region in the complex plane is given by the values of values of $\omega = \omega_x + i\omega_y$ that satisfy 
	\begin{align}
		\frac{1}{\sigma^2} &= \sum_k \gamma(k) \frac{(k/p)^2}{\omega_x^2[1 - \frac{\Gamma \sigma^2 k}{(1+\Gamma) p} h ]^2 + \omega_y^2[1 + \frac{\Gamma \sigma^2 k}{(1-\Gamma)p} h]^2}, \nonumber \\
		h &=  \sum_k \gamma(k) \frac{k/p}{\omega_x^2[1 - \frac{\Gamma \sigma^2 k}{(1+\Gamma)p} h]^2 + \omega_y^2[1 + \frac{\Gamma \sigma^2 k}{(1-\Gamma)p} h]^2} . \label{bulkboundary}
	\end{align}
	This is verified in the case of a dichotomous degree distribution in Fig. \ref{fig:modifiedellipse}. We now demonstrate that the elliptical law can be recovered from Eqs.~(\ref{bulkboundary}) in the case of an Erd\"os-Renyi graph \cite{erdos1960evolution}.
	
	One notes that for large $p$, the degree distribution of an Erd\"os-R\'enyi graph, which is Poissonian, tends towards a normal distribution with mean $p$ and variance $p$. Making the substitution $x = k/p$, we find that we can approximate the sums in Eqs.~(\ref{bulkboundary}) as integrals
	\begin{align}
		\frac{1}{\sigma^2} &= \int dx \frac{1}{\sqrt{2 \pi/p}} \frac{x^2  e^{-\frac{p(x-1)^2}{2}}}{\omega_x^2[1 - \frac{\Gamma \sigma^2 }{(1+\Gamma) } h  x]^2 + \omega_y^2[1 + \frac{\Gamma \sigma^2 }{(1-\Gamma)} h x]^2}, \nonumber \\
		h &=  \int dx \frac{1}{\sqrt{2 \pi/p}} \frac{x  e^{-\frac{p(x-1)^2}{2}}}{\omega_x^2[1 - \frac{\Gamma \sigma^2 }{(1+\Gamma)} h x]^2 + \omega_y^2[1 + \frac{\Gamma \sigma^2 }{(1-\Gamma)} h x]^2} . \label{bulkboundaryapprox}
	\end{align}
	In the limit $p\to \infty$, the Gaussian function in the integrand tends towards a Dirac delta function, i.e. $e^{-\frac{p(x-1)^2}{2}}/\sqrt{2 \pi/p} \to \delta(x-1)$. Hence, one finds $h = 1/\sigma^2$ by comparing both of Eqs.~(\ref{bulkboundaryapprox}). Subsequently, from the first of Eqs.~(\ref{bulkboundaryapprox}), one recovers the elliptic law \cite{sommers}, as expected 
	\begin{align}
		\frac{\omega_x^2}{(1+\Gamma)^2} + \frac{\omega_y^2}{(1-\Gamma)^2} = \sigma^2. \label{ellipticlaw}
	\end{align}
	This reasoning would also apply for any network with degree heterogeneity that vanishes as $p\to \infty$ (see Section \ref{section:corrections} for further discussion).
	
	\subsubsection{Leading eigenvalue of the bulk region}
	In many applications where one is concerned with determining the stability of a complex system, it is sufficient to deduce the location of the leading eigenvalue of the matrix $\underline{\underline{a}}$, rather than the full eigenvalue spectrum. Disregarding for now the outlier eigenvalue, the leading eigenvalue would be given by the location of the edge of the bulk region of the eigenvalue spectrum. 
	
	As can be seen by setting $\omega_y = 0$ and making the substitutions $h = (1+\Gamma) A/\lambda_{\mathrm{edge}}$ and $\omega_x = \lambda_{\mathrm{edge}}$ in Eqs.~(\ref{bulkboundary}), the leading eigenvalue of the bulk region is given by
	\begin{align}
		\frac{1}{\sigma^2} &= \sum_k \gamma(k) \frac{(k/p)^2}{(\lambda_{\mathrm{edge}} - \Gamma \sigma^2 k A/p)^2 } , \nonumber \\
		A &=  \sum_k \gamma(k) \frac{k/p }{(\lambda_{\mathrm{edge}} - \Gamma \sigma^2 k A/p) }. \label{leadingeigenvalue}
	\end{align}
	These equations can be solved simultaneously for $\lambda_{\mathrm{edge}}$, the extreme of the bulk region of the eigenvalue spectrum. The prediction of Eq.~(\ref{leadingeigenvalue}) is tested in the case of a uniform degree distribution in Figs. \ref{fig:bulkedgevsgamma} and \ref{fig:leadingeigenvalue}. 
	\subsubsection{Eigenvalue density in the bulk region}
\begin{figure*}[t]
		\centering 
		\includegraphics[scale = 0.52]{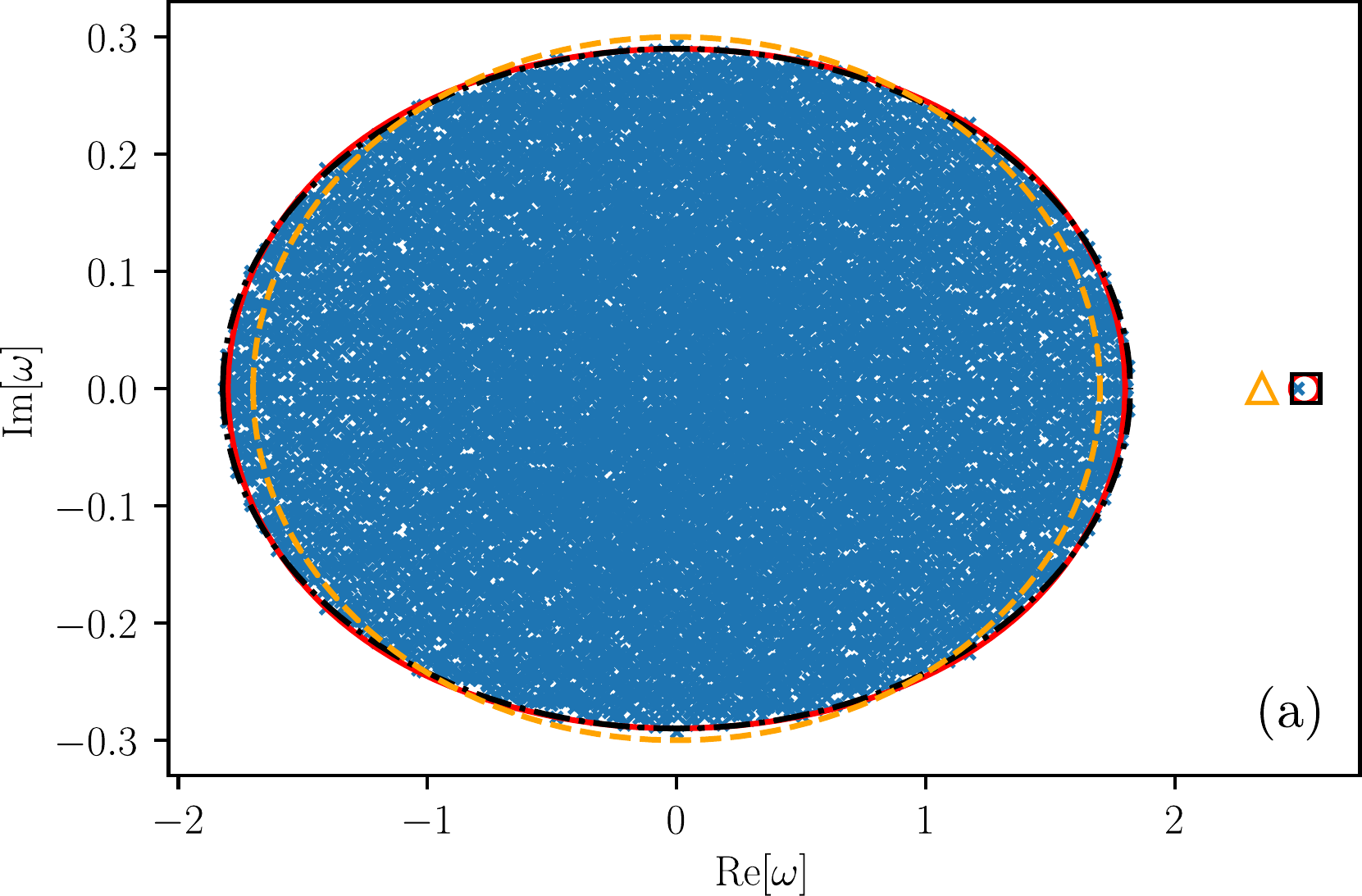}
		\includegraphics[scale = 0.52]{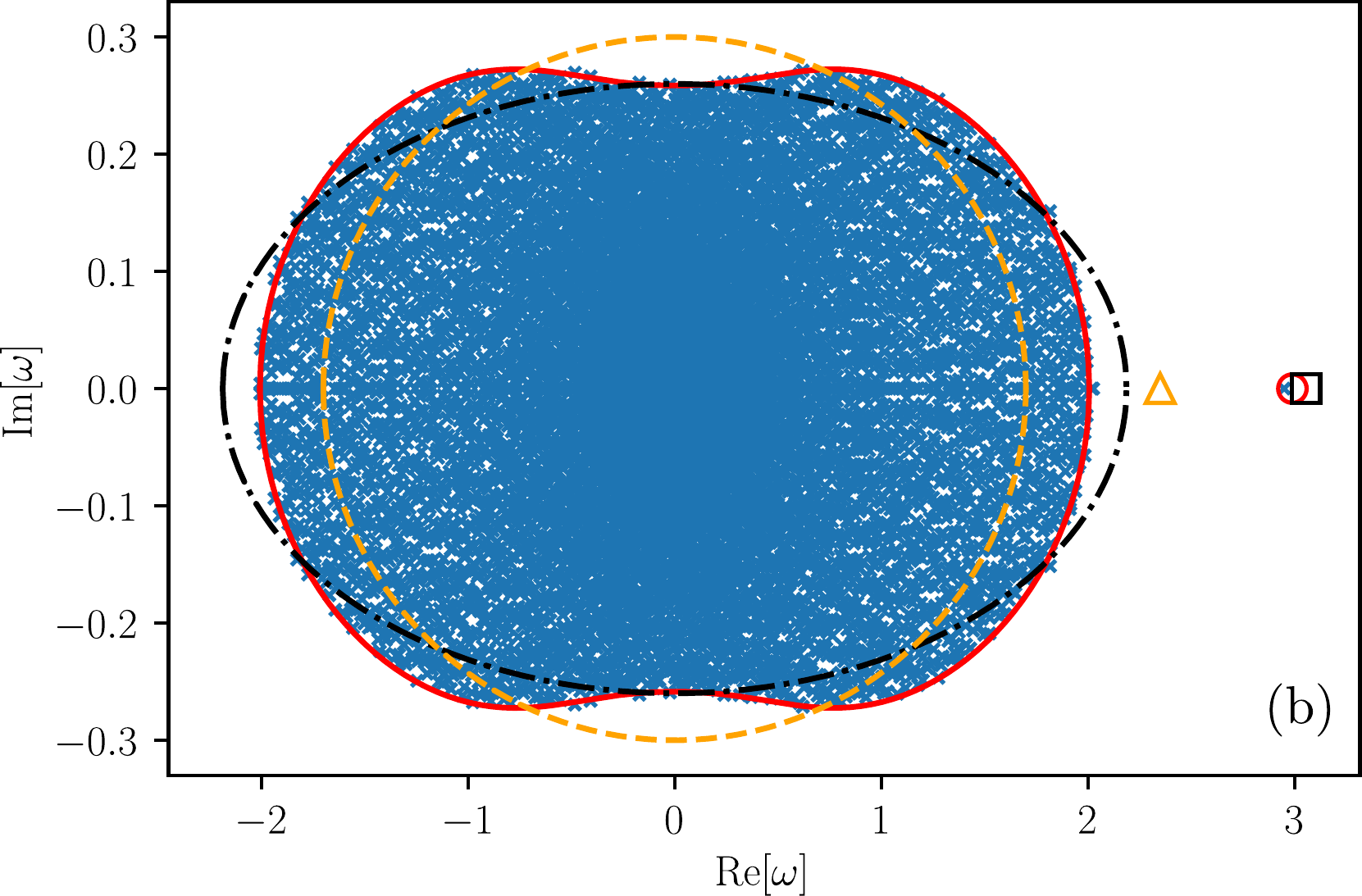}
		\caption{Validity of the small-heterogeneity approximation and verification of the precise results for the bulk boundary and outlier position. The results of numerical diagonalisation shown as blue crosses. In both panels, the degree distribution of the network is dichotomous $\gamma(k) = (\delta_{k, k_1}+ \delta_{k, k_2})/2$ with $\mu = 2$, $\sigma = 1$, $\Gamma = 0.7$, $N = 10000$ and $p = 200$. The na\"ive elliptical law and outlier [given in Eqs.~(\ref{ellipticlaw}) and (\ref{naiveoutlier}) respectively] that one would obtain for an Erd\"os-Renyi graph or a network with vanishing degree heterogeneity is shown as a dashed yellow line with yellow triangle indicating the outlier. The small-degree-heterogeneity approximation and the unapproximated theory predictions [from Eqs.~(\ref{modifiedellipse}) and (\ref{bulkboundary}) respectively] are shown as black dot-dashed and red solid lines respectively. The approximated [see Eq.~(\ref{modifiedoutlier})] and unapproximated [see Eq.~(\ref{outlier})] predictions for the outlier eigenvalue are shown as a black square and a red circle respectively. In panel (a) $k_1 = 145$ and $k_2 = 255$ giving $s^2 = 0.075$, and the approximate result is seen to be accurate. In panel (b) $k_1 = 90$ and $k_2 = 310$ giving $s^2 = 0.3$, and the small $s^2$ approximation fails. } \label{fig:modifiedellipse}
	\end{figure*}

	When the matrix $\underline{\underline{a}}$ corresponds to a fully-connected or Erd\"os-R\'enyi graph, the eigenvalue density inside the boundary of the bulk region (which is an ellipse in this case) is uniform \cite{sommers}. When the network structure is more complicated, the eigenvalue density can vary within the bulk region, which may also not be elliptical (see Fig. \ref{fig:modifiedellipse}).
	
	In SM Section S4 D, we show that one obtains for the trace of the resolvent matrix
	\begin{align}
		G(\omega, \omega^\star)&= \sum_k  \gamma(k) \frac{ \omega^\star - \Gamma \sigma^2 \frac{k}{p} \left(\frac{\omega - \Gamma \omega^\star}{1 - \Gamma^2} \right) m}{\left\vert \omega^\star - \Gamma \sigma^2 \frac{k}{p} \left(\frac{\omega - \Gamma \omega^\star}{1 - \Gamma^2} \right) m \right\vert^2 - \left(\frac{k}{p}\right)^2 g}, \label{gsum}
	\end{align}
	where we see that $G(\omega, \omega^\star)$ inside the bulk region of the eigenvalue spectrum is in general non-analytic. The functions $g(\omega, \omega^\star)$ and $m(\omega, \omega^\star)$ are given by the simultaneous solution of 
	\begin{align}
		\frac{1}{\sigma^2} &=\sum_{k} \gamma(k) \frac{(k/p)^2}{\left\vert \omega^\star - \Gamma \sigma^2 \frac{k}{p} \left(\frac{\omega - \Gamma \omega^\star}{1 - \Gamma^2} \right) m \right\vert^2 - \left(\frac{k}{p}\right)^2 g} , \nonumber \\
		m &= \sum_k \gamma(k) \frac{k/p}{\left\vert \omega^\star - \Gamma \sigma^2 \frac{k}{p} \left(\frac{\omega - \Gamma \omega^\star}{1 - \Gamma^2} \right) m \right\vert^2 - \left(\frac{k}{p}\right)^2 g} . \label{gsol}
	\end{align}
	Upon solving Eqs.~(\ref{gsol}) for $m(\omega,\omega^\star)$ and $g(\omega,\omega^\star)$, one substitutes these expressions into Eq.~(\ref{gsum}). One can then use Eq.~(\ref{bulkdensity}) to find the eigenvalue density.
	
	We now recover the usual uniform eigenvalue density for the elliptic law \cite{sommers}. In the case of an Erd\"os-Renyi graph with large $p$, we can approximate the degree distribution by $\gamma(k) = \delta_{k,p}$ [see the discussion surrounding Eq.~(\ref{bulkboundary})]. From Eqs.~(\ref{gsol}) we find $m  = 1/\sigma^2$. We thus find that $G(\omega, \omega^\star) = \frac{\omega^\star}{\sigma^2(1-\Gamma^2)} - \frac{\Gamma \omega }{\sigma^2(1-\Gamma^2)}$. Using Eq.~(\ref{bulkdensity}), we thus recover a uniform eigenvalue density inside the bulk region $\rho_{\mathrm{bulk}}(\omega, \omega^\star) = 1/[\pi \sigma^2(1 - \Gamma^2)]$.
	
	In general, it is impractical to find a full solution to Eqs.~(\ref{gsol}). However, as we demonstrate in Sections \ref{section:circlelaw} and \ref{section:banetwork}, it is possible to solve these equations explicitly in certain special cases and also to find informative approximations when the network heterogeneity is small.
	
	\subsection{Symmetric matrices}
	In the case where $\Gamma = 1$, the matrix $\underline{\underline{a}}$ becomes symmetric. This means that all the eigenvalues are real and the trace of the resolvent matrix $G$ is analytic (see Section S5 of the SM for a further discussion). When this is the case, we do not calculate a two-dimensional eigenvalue density in the complex plane, but instead the 1-dimensional density along the real axis. For this reason, Eq.~(\ref{bulkdensity}) is no longer applicable, and instead the eigenvalue density along the real axis can be found from the resolvent via \cite{edwardsjones}
	\begin{align}
		\rho_{x}(\omega_x) &= \frac{1}{\pi}\lim_{\epsilon \to 0}\mathrm{Im}\left[G(\omega_x - i \epsilon) \right]. \label{symmetricdensfromres}
	\end{align}
	The trace of the resolvent matrix is now given by (see Section S5 of the SM) 
	\begin{align}
		G(\omega_x) &= \sum_{k}\gamma(k) \frac{1}{\omega_x - \sigma^2 (k/p) A}, \nonumber \\
		A &= \sum_k \gamma(k) \frac{k/p}{\omega_x - \sigma^2 (k/p) A} . \label{symmetricmatrixdensity}
	\end{align}
	Now, we once again examine the case $\gamma(k) = \delta_{k, p}$, which is equivalent to studying an Erd\"os-R\'enyi graph in the limit $p\to\infty$. From Eq.~(\ref{symmetricmatrixdensity}), we obtain $A = G = [\omega_x - \sqrt{\omega_x^2- 4 \sigma^2}]/(2 \sigma^2)$. Substituting this into Eq.~(\ref{symmetricdensfromres}), one recovers the Wigner semi-circle law \cite{wigner1958distribution, wigner1967random} 
	\begin{align}
		\rho_{x}(\omega_x) = \frac{1}{2\pi\sigma^2} \sqrt{4 \sigma^2 - \omega_x^2}, \label{wignersemicircle}
	\end{align}
	where $\rho_x(\omega_x) = 0$ for $\omega_x^2>4 \sigma^2$.

	\subsection{Outlier eigenvalues}\label{section:outliers}
	When $\mu = 0$ [see Eq.~(\ref{matrixensemble})], all of the eigenvalues of the matrix $\underline{\underline{a}}$ are confined to the bulk region of the complex plane. However, when a non-zero value of $\mu$ is introduced, a single outlier eigenvalue may protrude from the bulk region. 
	
	The observation that is key to deducing the outlier eigenvalue is that, because we can use the annealed approximation for the statistics of the elements $a_{ij}$ [given in Eq.~(\ref{annealedstatistics})], the introduction of a non-zero value of $\mu$ is akin to making a rank-1 perturbation to the matrix with $\mu = 0$. 
	
	Introducing $(\underline{\underline{\mu}})_{ij} = \mu k_i k_j/(N) $, and letting $\underline{\underline{z}} = \underline{\underline{a}} - p^{-1}\underline{\underline{\mu}}$, any eigenvalue $\lambda$ of the random matrix $\underline{\underline{a}}$ must satisfy (by definition)
	\begin{align}
		\det\left[ \lambda\underline{\underline{\id}} - \underline{\underline{z}} - p^{-1}\underline{\underline{\mu}}\right] = 0.
	\end{align}
	If $\lambda = \lambda_{\mathrm{outlier}}$ is outside the bulk of the eigenvalue spectrum, the matrix $\lambda_{\mathrm{outlier}}\underline{\underline{\id}} - \underline{\underline{z}}$ is invertible, and hence 
	\begin{align}
		\det\left[ \underline{\underline{\id}} - p^{-1}\underline{\underline{G_0}}\,\underline{\underline{\mu}}\right] = 0,
	\end{align}
	where $\underline{\underline{G_0}}(\lambda_{\mathrm{outlier}}) =  \left[\lambda_{\mathrm{outlier}} \underline{\underline{\id}} - \underline{\underline{z}}\right]^{-1}$ is the resolvent matrix of $\underline{\underline{a}}$ when $\mu = 0$ [see Eq.~(\ref{resolventdef})], which is available to us from the calculation of the bulk spectrum.
	
	\begin{figure*}[t]
		\centering 
		\includegraphics[scale = 0.55]{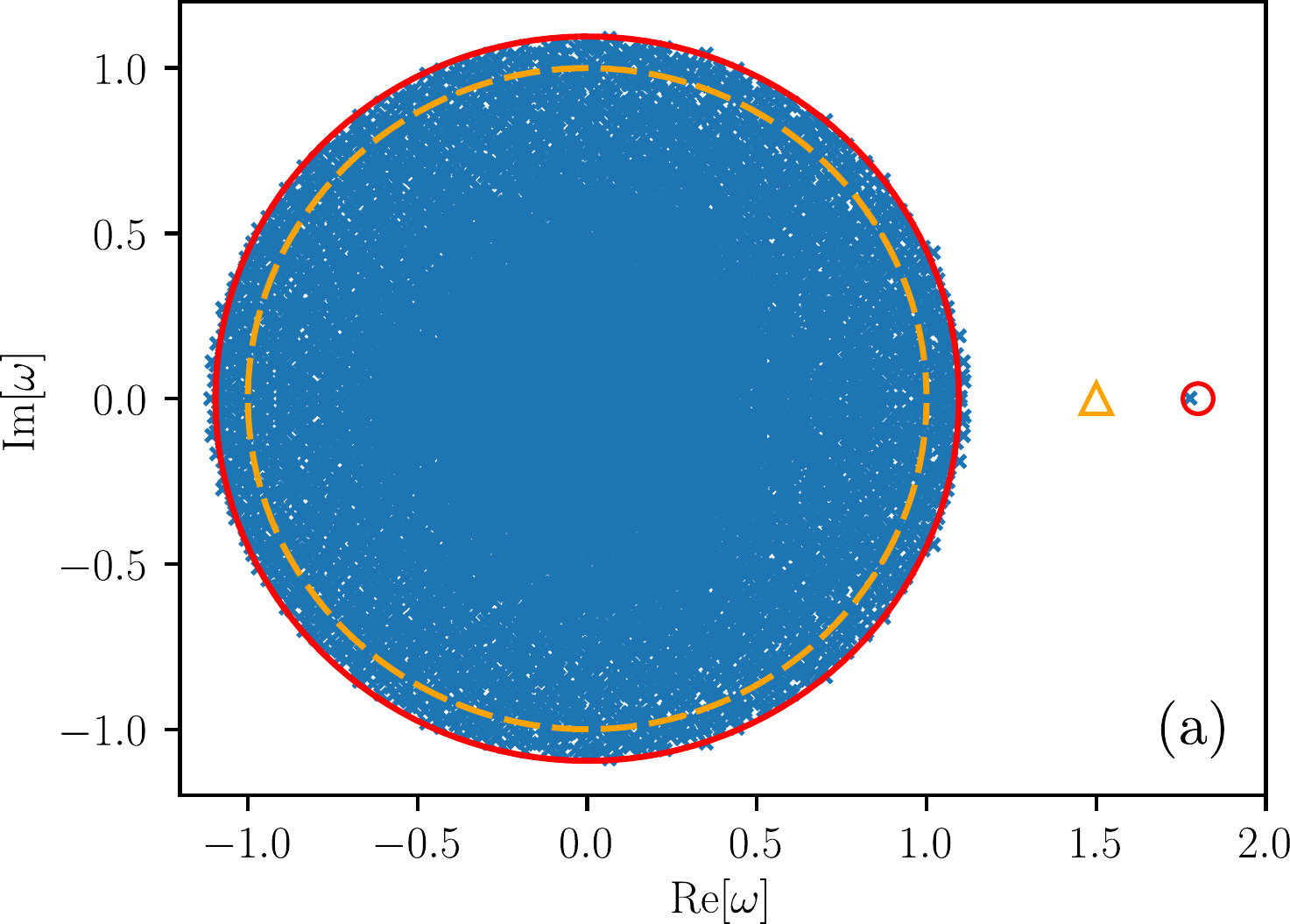}
		\includegraphics[scale = 0.55]{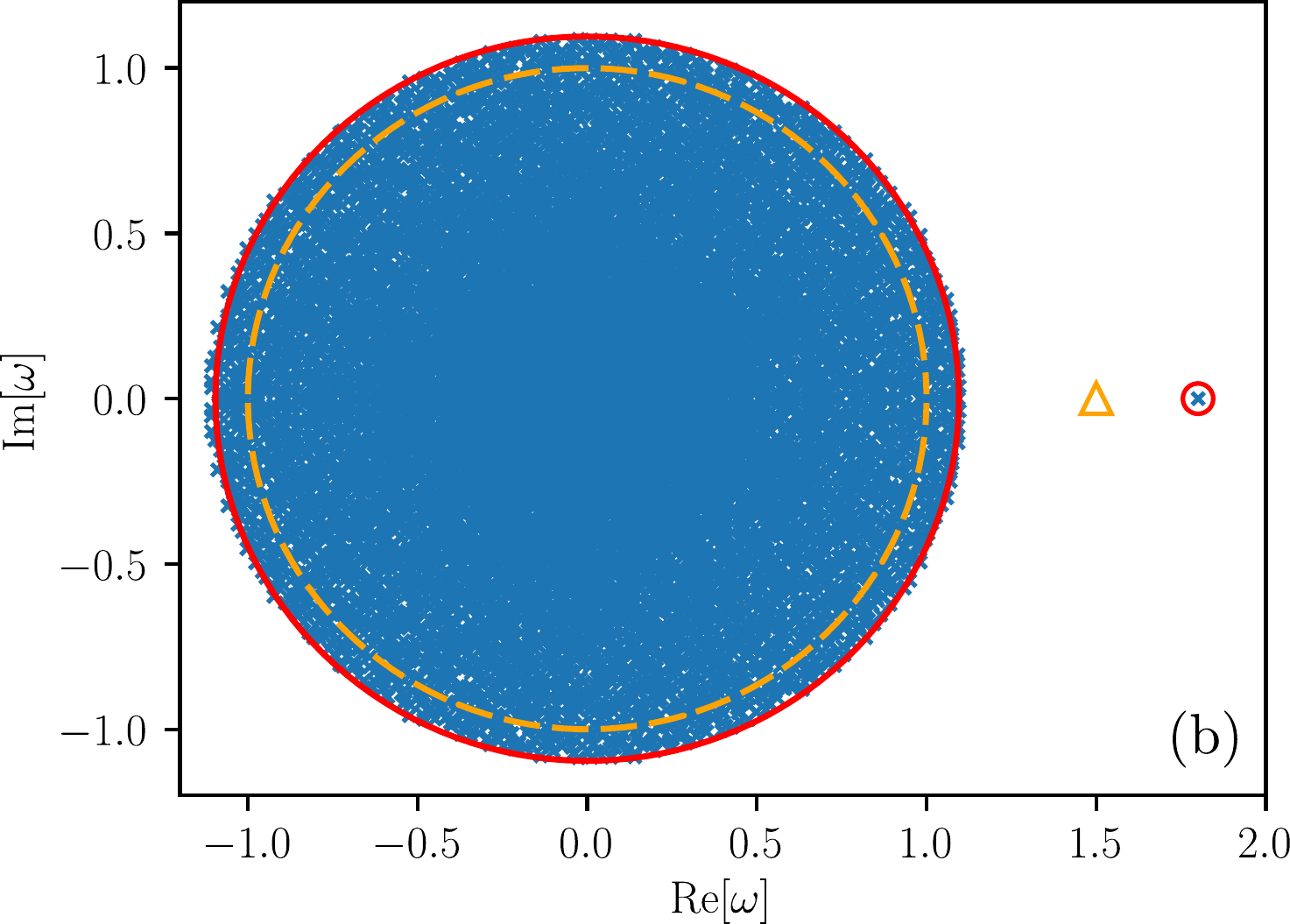}
		\caption{Verification of the universal circle and outlier laws for $\Gamma = 0$ given in Eqs.~(\ref{universalcircle}) and (\ref{universaloutlier}) [solid red lines and circular points respectively]. The prediction in the absence of degree heterogeneity ($s^2 = 0$) for the bulk region boundary and the corresponding outlier are represented by a dashed orange line and an orange triangle. In both panels, $\mu = 1.5$, $\sigma = 1$, $\Gamma = 0$, $N = 10000$, $p = 200$ and $s^2 = 0.2$. In panel (a), the dichotomous distribution from Fig. \ref{fig:modifiedellipse} was used but with $k_1 = 111$ and $k2 = 289$, whereas in panel (b) the uniform distribution described in Fig. \ref{fig:circlelawdensity} was used (but with $s^2 = 0.2$).} \label{fig:universality}
	\end{figure*}
	
	Finally, by using Sylvester's determinant identity \cite{orourke, baron2020dispersal} (see Section S6 of the SM for further details), we obtain a general expression for the outlier eigenvalue
	\begin{align}
		\frac{1}{\mu} &=\sum_{k}\gamma(k) \frac{(k/p)^2}{\lambda_{\mathrm{outlier}} - \Gamma \sigma^2 k A/p } , \nonumber \\
		A &= \sum_k \gamma(k)\frac{k/p }{\lambda_{\mathrm{outlier}} - \Gamma \sigma^2 k A/p } . \label{outlier}
	\end{align}
	Eqs.~(\ref{outlier}) are verified in Figs.~\ref{fig:modifiedellipse} and \ref{fig:leadingeigenvalue} in the case of a dichotomous and a uniform degree distribution respectively. 
	
	In the case of an Erd\"os-R\'enyi graph for large $p$, for which $\gamma(k) \approx \delta_{k,p}$, we find from Eqs.~(\ref{outlier}) that $A =1/\mu$. Substituting this result back into the first of Eqs.~(\ref{outlier}), we obtain the known result \cite{orourke, baron2022eigenvalues}
	\begin{align}
		\lambda_{\mathrm{outlier}} = \mu + \frac{\Gamma \sigma^2}{\mu}. \label{naiveoutlier}
	\end{align}

	\pagebreak
	\section{Corrections to known results due to non-zero network degree heterogeneity}\label{section:corrections}
	In this section, we use the formulae presented in Section \ref{section:generalresults} to derive simpler closed-form approximations for the eigenvalue spectrum. We derive corrections to the usual elliptic, circular and semi-circular laws in the case of non-vanishing degree heterogeneity in the limit $p\to \infty$, as well as the outlier eigenvalues. 
	
	For the remainder of this work, we define the degree heterogeneity of the network by
	\begin{align}
		s^2 = \sum_k \gamma(k) (k-p)^2/p^2 .
	\end{align}
	For the Erd\"os-Renyi graph, we have $s^2 = 1/p$, which vanishes in the limit $p\to\infty$. We will see in this section that all graphs with vanishing degree heterogeneity in the limit $p\to\infty$ share the same eigenvalue spectrum as the Erd\"os-Renyi graph in this limit. 
	
	\subsection{Universal circular law and outlier}\label{section:circlelaw}

	We first examine the case where $\Gamma = 0$. From Eq.~(\ref{bulkboundary}), we obtain the rather elegant result that the boundary of the bulk of the eigenvalue spectrum is given by a modified circular law that takes into account the heterogeneity of the network
	\begin{align}
		\omega_x^2 + \omega_y^2 = \sigma^2 (1 + s^2).\label{universalcircle}
	\end{align}
	We note that this result is entirely independent of the precise degree distribution $\gamma(k)$, depending only on its second moment. In this sense, it is universal \cite{taovukrishnapur2010, taovu2010}. We thus see that as a result of network degree heterogeneity, the eigenvalues become more broadly distributed. 
	
	The result in Eq.~(\ref{universalcircle}) is very similar to one derived recently by Neri and Metz \cite{neri2020linear}. This work deals with graphs where nodes are allowed to have differing `in'- and `out'- degrees. In Ref. \cite{neri2020linear}, $s^2$ in Eq.~(\ref{universalcircle}) is replaced by the `degree correlation coefficient', which quantifies the extent to which the `in'- and `out'- degrees of nodes on the network are related. 
	
	From Eq.~(\ref{outlier}), we also find in the case $\Gamma = 0$ that 
	\begin{align}
		\lambda_{\mathrm{outlier}} = \mu (1 + s^2).\label{universaloutlier}
	\end{align}
	Similarly, this result is independent of the precise degree distribution of the network. The results in Eqs.~(\ref{universaloutlier}) and (\ref{universalcircle}) are tested in Fig. \ref{fig:universality} for both uniform and dichotomous degree distributions.
	
	From Eqs.~(\ref{universalcircle}) and (\ref{universaloutlier}), we can deduce that for $\Gamma = 0$ network heterogeneity is a purely destabilising influence (see the Section \ref{section:stability} for further comment). We note that this is not necessarily the case for other values of $\Gamma$, as is shown in the following subsections. 
	
	However, despite the universality of the preceding results, the density of the eigenvalues inside the circular bulk region is not universal in the same way. One can show from Eqs.~(\ref{gsum}) -- (\ref{gsol}) (see SM Section S7 A1) that
	
	\begin{align}
		\rho(\vert\omega\vert) &= \frac{1}{\pi} \mathrm{Re}\left\{\frac{\partial }{\partial \omega^\star} \left[\sum_{k}\gamma(k)  \frac{\omega^\star}{\vert \omega\vert^2 - (k/p)^2 g(\vert \omega\vert)}\right] \right\}\nonumber \\
		&= \frac{1}{\pi} \mathrm{Re} \left\{ h_{0,1} + \vert \omega\vert^2 [(h_{2,2})^2/h_{4,2} - h_{0,2}]\right\} , \label{circledensity}
	\end{align}
	where we have defined the functions
	\begin{align}
		h_{n,m}(\vert\omega\vert) &\equiv \sum_k \gamma(k)  \frac{(k/p)^n}{\left[\vert \omega\vert^2 -(k/p)^2 g(\vert \omega\vert)\right]^m },
	\end{align}
	and where $g(\vert \omega\vert)$ is determined by
	\begin{align}
		\frac{1}{\sigma^2} = h_{2,1}(g, \vert\omega\vert) . \label{gcond}
	\end{align}
	
	\begin{figure}[t]
		\centering 
		\includegraphics[scale = 0.5]{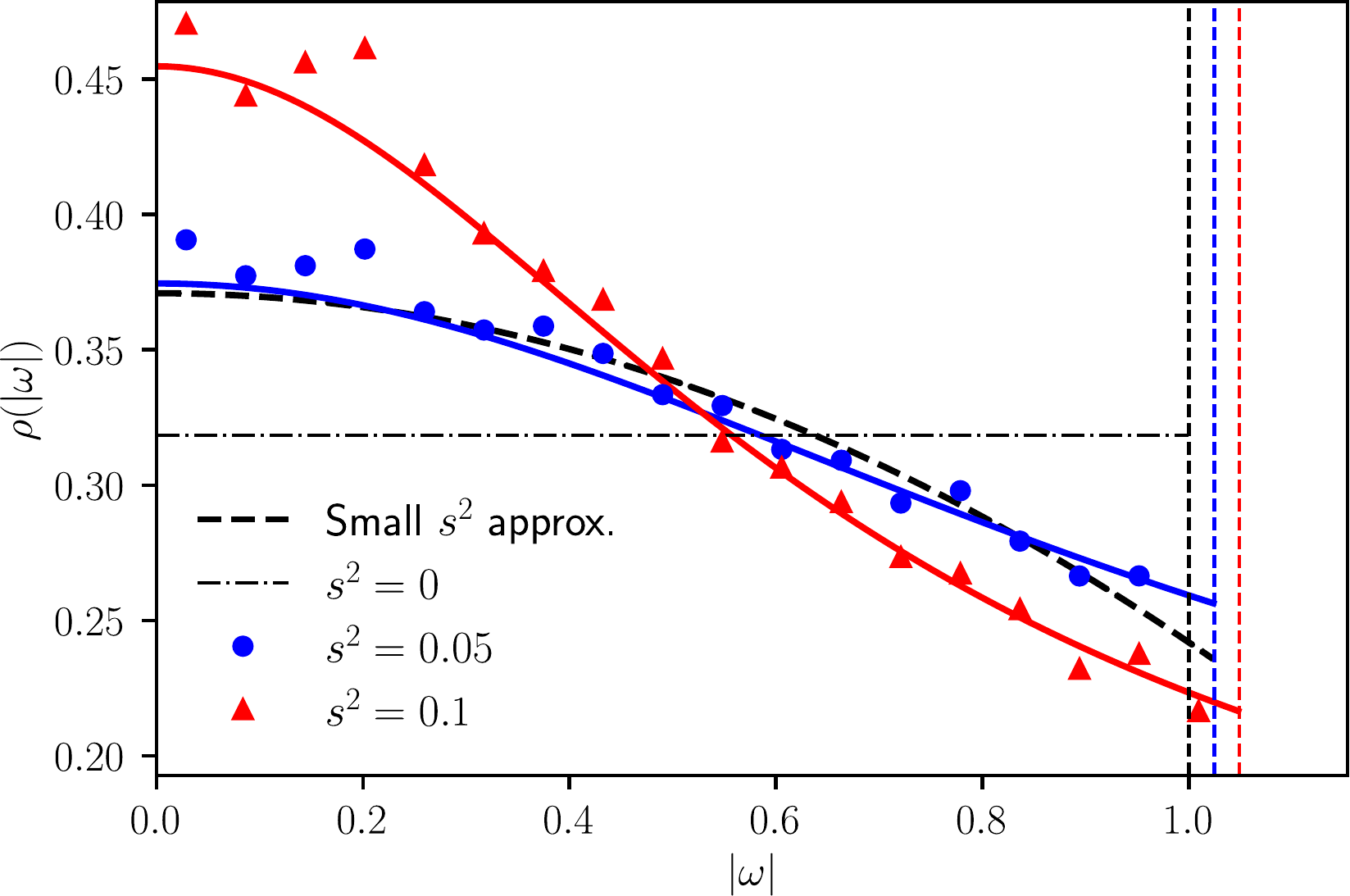}
		\caption{Modified eigenvalue density inside the circular bulk region due to non-zero network degree heterogeneity. The degree distribution is uniform: $\gamma(k) = \sum_{l = p - [\sqrt{3} s p]}^{p +[\sqrt{3} s p] }\delta_{k,l}/(2[\sqrt{3} s p])$. Markers are the results of the numerical diagonalisation of a single random matrix. Model parameters are $\mu = 0$, $p = 200$, $N = 10000$, $\Gamma = 0$ and $\sigma = 1$. The horizontal dot-dashed line shows the uniform value that the eigenvalue density would take without network heterogeneity ($s^2 = 0$). Solid lines show the theory predictions from Eqs.~(\ref{circledensity}). The dashed curve is the small $s^2$ approximation for $s^2 = 0.05$ given in Eq.~(\ref{densityapprox}). Vertical dashed lines show the position of the edge of the eigenvalue spectrum for each value of $s^2$.} \label{fig:circlelawdensity}
	\end{figure}
	
	Eqs.~(\ref{circledensity})--(\ref{gcond}) are clearly dependent on the precise choice of $\gamma(k)$. However, for small values of $s^2$, we show (again in SM Section S7 A1) that the eigenvalue density inside the circular bulk region can be approximated by
	\begin{align}
		\rho(\vert \omega\vert) = \frac{1}{\pi \sigma^2(1 - 5 s^4 - 4 s^6)} \left[ 1 + s^2 \left(3 - 8\frac{\vert \omega\vert^2}{\sigma^2}\right) \right] + O(s^4). \label{densityapprox}
	\end{align}
	One notes that as a result of degree heterogeneity, the eigenvalues become more concentrated at the centre of the circle (i.e. the density is a decreasing function of $\vert\omega\vert$). The expression in Eq.~(\ref{densityapprox}) is valid for any degree distribution, as long as $s^2$ is small. The expressions in Eqs.~(\ref{circledensity}) -- (\ref{densityapprox}) are verified in Fig. \ref{fig:circlelawdensity}.
	
	We note that for $s^2\to 0$, Eqs.~(\ref{universalcircle}), (\ref{universaloutlier}) and (\ref{densityapprox}) reduce to known results that are valid for networks without degree heterogeneity \cite{taobook} (which are given in Section \ref{section:generalresults}).
	
	\medskip
	
	\subsection{Modified elliptic law and outlier for small degree heterogeneity}
	\begin{figure}[t]
		\centering 
		\includegraphics[scale = 0.53]{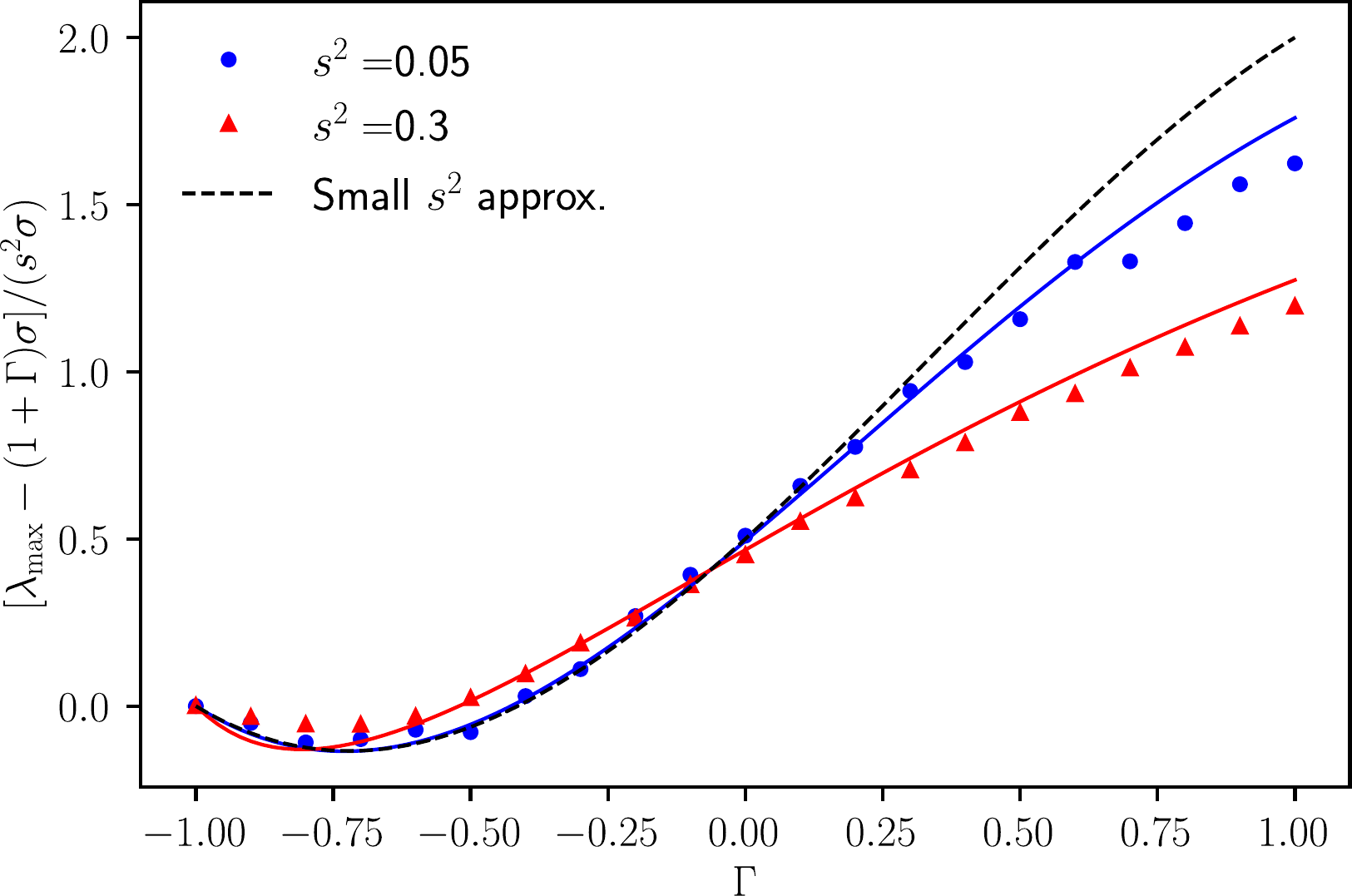}
		\caption{Verification of  the first-order correction to the leading eigenvalue of the bulk region of the eigenvalue spectrum. The prediction for small degree heterogeneity, $[\lambda_{\mathrm{edge}} - (1+\Gamma) \sigma]/(s^2 \sigma) = 1/2 (1+\Gamma) (1 + 2 \Gamma - \Gamma^2)$ from Eq.~(\ref{bulkedge}) is shown as a dashed line. This is independent of $s^2$. Exact predictions in the cases $s^2 = 0.05$ and $s^2 = 0.3$, derived from Eq.~(\ref{leadingeigenvalue}), are shown as solid lines. The results of numerical diagonalisation are shown as markers. The remaining model parameters were $\mu = 0$, $p/N = 0.2$, $N = 12000$, and $\sigma = 1$, and the degree distribution was uniform, as specified in Fig. \ref{fig:circlelawdensity}. As $s^2$ is reduced, we see better agreement with the approximate result. } \label{fig:bulkedgevsgamma}
	\end{figure}
	
	In this section, we aim to find general expressions, analogous to Eqs.~(\ref{universalcircle}) and (\ref{universaloutlier}), in the case where $\Gamma \neq 0$. In this more general setting, we are unable derive such a simple universal law as for the special case $\Gamma = 0$. However, for small $s^2$, we can still derive informative and useful approximations. Specifically, we are able to understand how the usual elliptic law \cite{sommers} and its corresponding outlier \cite{baron2022eigenvalues, orourke} are modified by the introduction of degree heterogeneity.
	
	Beginning with Eqs.~(\ref{bulkboundary}) and expanding in small $s^2$, one finds that to leading order, the bulk of the eigenvalue spectrum falls within an ellipse, similar to the $s^2 = 0$ case \cite{sommers}. However, the major and minor axes of the ellipse are modified by $s^2>0$. That is, one finds that the majority of the eigenvalues fall within a boundary in the complex plane given by (see SM Section S7 A2)	
	
	\begin{align}
		&\frac{\omega_x^2}{(1 + \Gamma)^2 [1 + s^2/2 (1 + 2 \Gamma - \Gamma^2)]^2} \nonumber \\
		&+ \frac{\omega_y^2}{(1 - \Gamma)^2 [1 + s^2/2 (1 - 2 \Gamma - \Gamma^2)]^2} = \sigma^2. \label{modifiedellipse}
	\end{align} 
	The modified elliptic law in Eq.~(\ref{modifiedellipse}) is verified in Fig. \ref{fig:modifiedellipse}a.
	
\begin{figure}[t]
		\centering 
		\includegraphics[scale = 0.53]{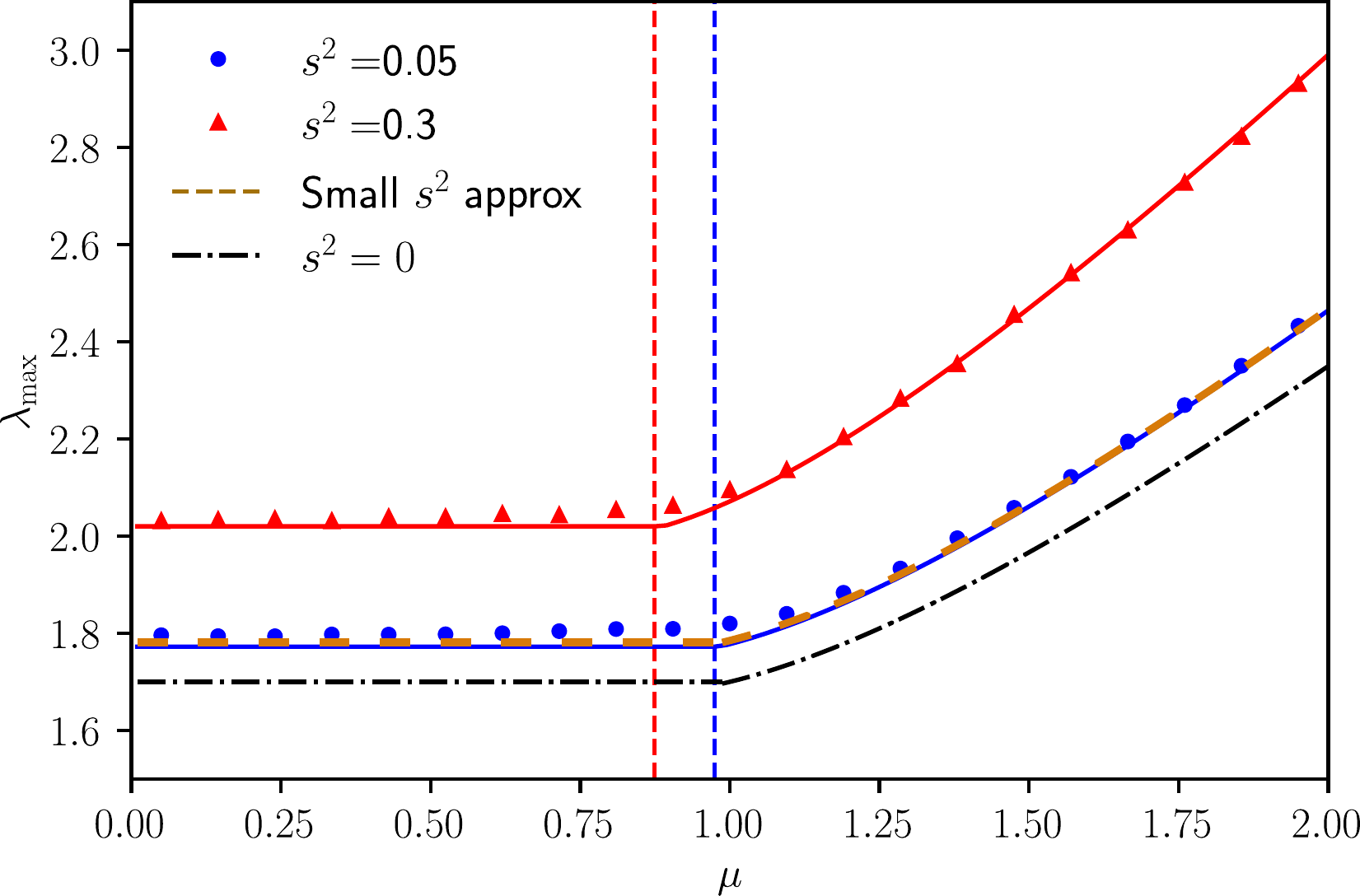}
		\caption{Leading eigenvalue as a function of $\mu$ for two values of $s^2$. The small $s^2$ approximation for the bulk edge in Eq.~(\ref{bulkedge}) and the outlier in Eq.~(\ref{modifiedoutlier}) are shown for $s^2 = 0.05$. Vertical dashed lines show the values of $\mu$ at which the outlier emerges from the bulk region. The solid lines show the unapproximated theory from Eqs.~(\ref{leadingeigenvalue}) and (\ref{outlier}). The black dot-dashed curve shows the prediction that doesn't take into account network degree heterogeneity, given in Eqs.(\ref{ellipticlaw}) and (\ref{naiveoutlier}). The remaining model parameters are $\Gamma = 0.7$, $p/N = 0.2$, $N = 12000$, and $\sigma = 1$, and the degree distribution is uniform, as specified in Fig. \ref{fig:circlelawdensity}.}\label{fig:leadingeigenvalue}
	\end{figure}

	The leading eigenvalue can also be derived by expanding Eqs.~(\ref{leadingeigenvalue}). The small $s^2$ approximation to the leading edge of the bulk of the eigenvalue spectrum is 
	\begin{align}
		\lambda_{\mathrm{edge}} = (1 +\Gamma)[1 + s^2/2 (1 + 2 \Gamma - \Gamma^2)] + O(s^4), \label{bulkedge}
	\end{align}
	which is in agreement with Eq.~(\ref{modifiedellipse}) to first order in $s^2$. This is verified in Figs. \ref{fig:bulkedgevsgamma} and \ref{fig:leadingeigenvalue}. 
	
	We see from Eq.~(\ref{bulkedge}) and Fig. \ref{fig:bulkedgevsgamma} that degree heterogeneity can have the effect of both stabilising or destabilising a dynamical system depending on the value of $\Gamma$ (see Section \ref{section:stability} for further discussion). The critical value at which the effect of degree heterogeneity switches from being stabilising to destabilising (for small $s^2$) is $\Gamma~=~1~-~\sqrt{2}$. 
	
	In the case where $\Gamma \neq 0$, we can also find a small $s^2$ approximation to the outlier eigenvalue. Expanding Eqs.~(\ref{outlier}) for small $s^2$ (see SM Section S7 A2), we find the surprisingly simple formula
	\begin{align}
		\lambda_{\mathrm{outlier}} = (1 + s^2)\left(\mu + \frac{\Gamma \sigma^2}{\mu}\right) + O(s^4) . \label{modifiedoutlier}
	\end{align}
	Here, we see that degree heterogeneity acts to further remove the outlier from the bulk region of the eigenvalue spectrum, making network degree heterogeneity a purely destabilising influence in cases where $\lambda_{\mathrm{outlier}}$ is the leading eigenvalue. Eq.~(\ref{modifiedoutlier}) is verified in Fig.\ref{fig:leadingeigenvalue}.
	
	Both Eq.~(\ref{modifiedellipse}) and Eq.~(\ref{modifiedoutlier}) can easily be seen to reduce to Eqs.~(\ref{ellipticlaw}) and (\ref{naiveoutlier}) respectively in the case $s^2 \to 0$. This demonstrates that any network with vanishing degree heterogeneity in the limit $p\to \infty$ satisfies the usual elliptic/outlier law \cite{sommers, orourke}.

	\subsection{Undirected networks: modified semi-circular law}
	In the case where $\Gamma = 1$ and the matrix $\underline{\underline{a}}$ is consequently symmetric, we can also derive a correction to the well-known Wigner semi-circle law \cite{wigner1958distribution, wigner1967random} for the case of non-zero network degree heterogeneity. One can show (see SM Section S7 A3) that the eigenvalue density along the real axis in Eq.~(\ref{symmetricmatrixdensity}) can be approximated by the following formula for small $s^2$
	\begin{align}
		\rho(\omega_x) &= \frac{2}{\pi \omega_c^2} (\omega_c^2 - \omega_x^2)^{1/2}\left[ 1 + 2 s^2 \left( 1 - \frac{4 \omega_x^2}{\omega_c^2}\right) \right] + O(s^4), \nonumber \\
		\omega_c^2 &= 4 \sigma^2 (1 + 2 s^2) + O(s^4) , \label{modifiedsemicircle}
	\end{align}
	where $\omega_c$ is the critical value of $\omega_x$ at which the eigenvalue density becomes zero, which agrees with the formula in Eq.~(\ref{bulkedge}) (to leading order in $s^2$). We note that Eq.~(\ref{modifiedsemicircle}) reduces to the familiar semi-circle law [see Eq.~(\ref{wignersemicircle})] in the limit $s^2 \to 0$. Further, we also note the resemblance between this formula and the approximation of the eigenvalue spectrum of symmetric sparse random matrices with Erd\"os-R\'enyi network structure \cite{rodgers1988density}, where here we have expanded in $s^2$, rather than $1/p$.
	
\begin{figure}[t]
		\centering 
		\includegraphics[scale = 0.53]{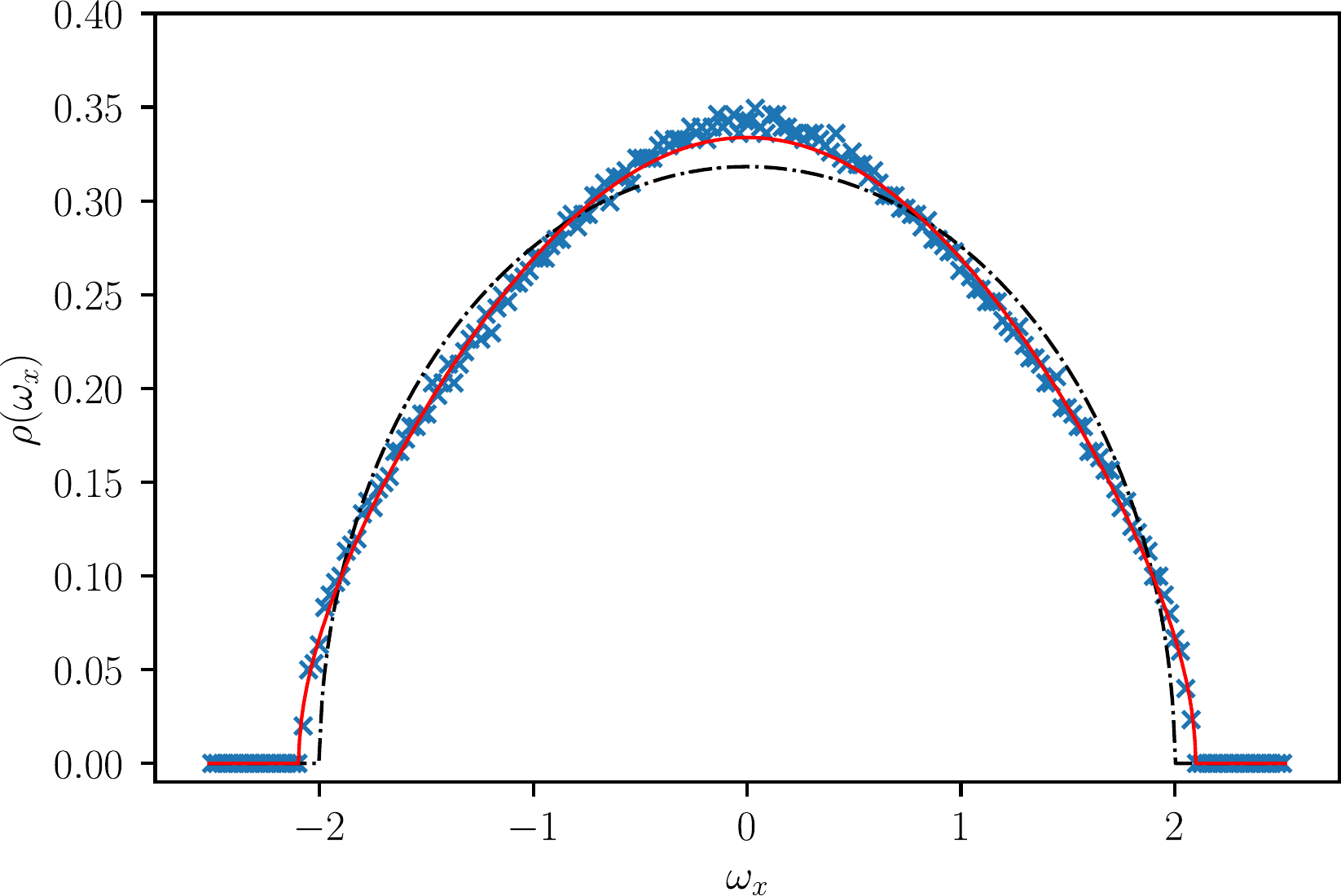}
		\caption{Modified semi-circle law due to network degree heterogeneity. The solid red line shows the prediction in Eq.~(\ref{modifiedsemicircle}), the dot-dashed black line shows the Wigner semi-circle law in Eq.~(\ref{wignersemicircle}). The blue crosses are the results of the a single diagonalisation of a random matrix with $N = 12000$, $p/N = 0.02$, $\mu = 0$, $\sigma = 1$, $\Gamma = 1$ and $s^2 = 0.05$ with the degree distribution of the network being uniform as described in the caption of Fig. \ref{fig:circlelawdensity}. } \label{fig:semicircle}
	\end{figure}

	The modified semi-circle law in Eq.~(\ref{modifiedsemicircle}) is verified in Fig. \ref{fig:semicircle}. 
	
	\pagebreak
	\section{Diverging degree heterogeneity: Barabasi-Albert networks}\label{section:banetwork}
	
	In this section, we demonstrate how the approach presented here can be used to produce closed-form results for the eigenvalue spectrum of a non-trivial directed complex network without a small $s^2$ approximation. We take as our example the paradigmatic Barabasi-Albert (BA) network. For simplicity, we take the case where $\Gamma = 0$. 
	
	The BA network \cite{albert2002statistical}, is constructed by a preferential attachment process. Beginning with $p$ fully-connected nodes (where $p$ is an even number), one attaches one new node at a time to $p/2$ of the nodes that make up the current network. Each node in the network is chosen with a probability proportional to its current degree. As a result of this process, the degree distribution of the network is given by (for large $N$) \cite{albert2002statistical}
	\begin{align}
		\gamma(k) = \frac{p (p+2)}{2 k (k+1)(k+2)}. \label{badegreedist}
	\end{align}
	The degree heterogeneity of the BA network diverges logarithmically with the network size $N$ \cite{vazquez2008analytical}. Because of this, in the limit $N\to \infty$, the eigenvalue spectrum of the Barabasi-Albert graph covers the entire complex plane [as can be seen from Eq.~(\ref{universalcircle})]. Despite this, we can still compute its eigenvalue density. 
	
\begin{figure}[H]
		\centering 
		\includegraphics[scale = 0.5]{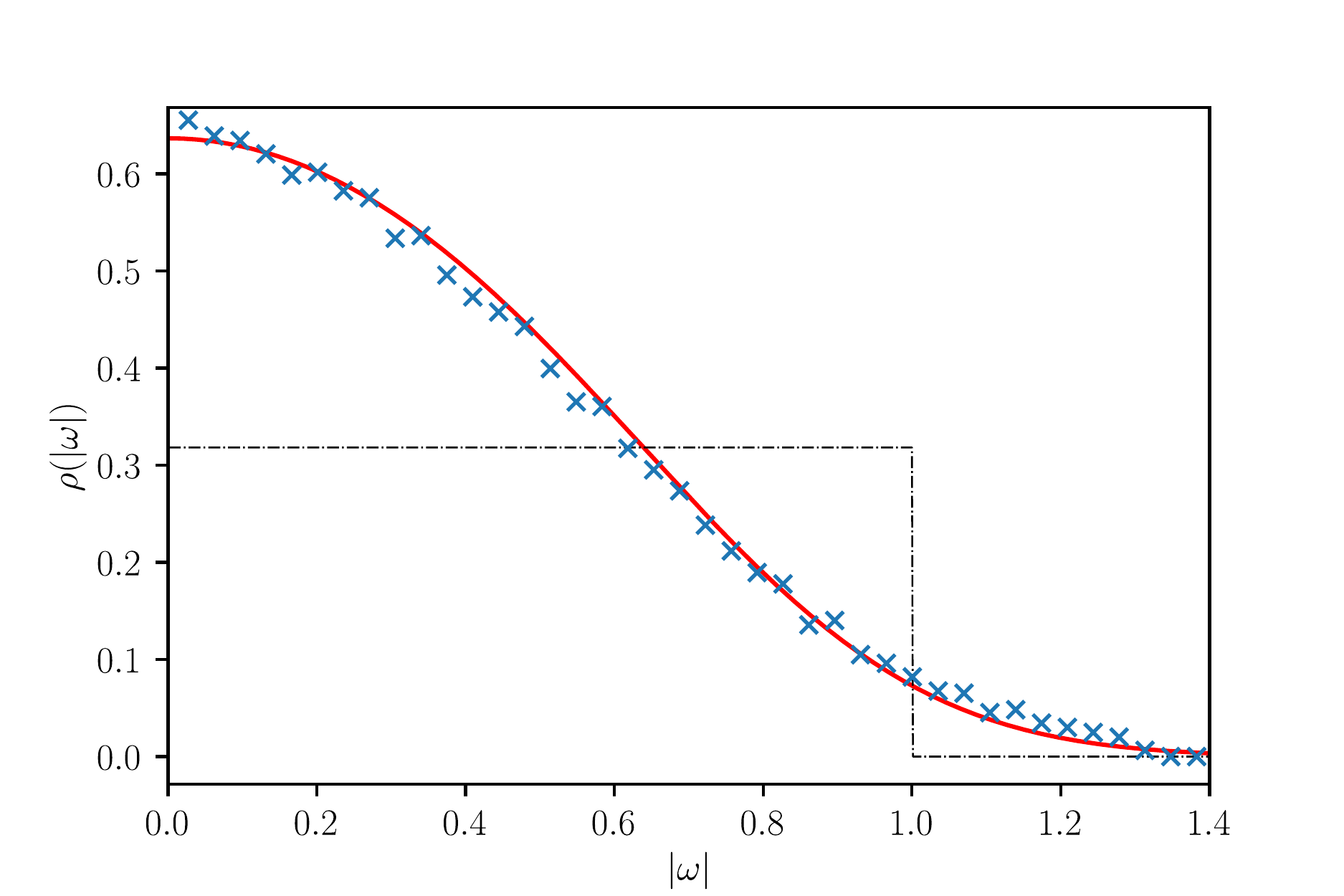}
		\caption{The eigenvalue spectrum of a Barabasi-Albert network with $\Gamma = 0$, $\mu = 0$, $\sigma = 1$, $N = 12000$, $p/N= 0.02$. Points are the result of diagonalising a single random matrix. The red line is the prediction in Eq.~(\ref{badensity}). The dot-dashed line is what one would expect from a network with vanishing degree heterogeneity (i.e. the usual circle law). } \label{fig:banetwork}
	\end{figure}

	We begin by solving Eq.~(\ref{gcond}) for $g(\vert \omega\vert)$. We have
	\begin{align}
		\sum_{k = p/2}^{\infty} \frac{p (p+2)}{2 k (k+1)(k+2)} \frac{(k/p)^2}{\vert \omega\vert^2 - (k/p)^2 g(\vert \omega \vert)} = \frac{1}{\sigma^2}.
	\end{align}
	Making the substitution $x = (2k/p)^{-2}$ and assuming that $p$ is large, the above sum can be approximated by the following integral
	\begin{align}
		\frac{1}{\sigma^2} \approx \int_{0}^{1} dx  \frac{1/4}{x\vert \omega\vert^2 - g(\vert \omega \vert)/4} = \frac{1}{4 \vert \omega\vert^2}\ln\left[1 - \frac{4 \vert \omega\vert^2}{g} \right],
	\end{align}
	which can be solved readily for $g(\omega)$. Once we have $g(\vert \omega\vert)$, we can evaluate the expression in Eq.~(\ref{circledensity}) for the eigenvalue density (approximating the sum as an integral in a similar fashion), and we obtain
	\begin{align}
		\rho(\omega) &= \frac{1}{\pi} \mathrm{Re}\left\{\frac{\partial}{\partial \omega^\star} \left[\int_0^1 dx \frac{\omega^\star x}{x\vert \omega\vert^2 - g(\vert \omega \vert)/4} \right]  \right\} \nonumber \\
		&= \frac{4\left[1 +e^{4\vert \omega\vert^2/\sigma^2} \left(\frac{4\vert\omega\vert^2}{\sigma^2} -1 \right)\right]}{\pi \sigma^2 \left( e^{4\vert \omega\vert^2/\sigma^2}-1\right)^2}. \label{badensity}
	\end{align}
	This expression for the eigenvalue density is verified in Fig. \ref{fig:banetwork}. We see that the eigenvalue density decays exponentially with $\vert\omega\vert^2$ as $\vert \omega\vert \to\infty$. This is in contrast to the eigenvalue spectrum of an undirected scale-free network [i.e. with $\Gamma = 1$, but the same $k^{-3}$ scaling behaviour as in Eq.~(\ref{badegreedist})], which was dealt with in Ref. \cite{rodgers2005eigenvalue}. In that case, it was found that the tails of the eigenvalue distribution scaled as $\rho(\vert\omega\vert) \sim \vert \omega\vert^{-5}$.
	
	\section{Applications: The effect of network heterogeneity on stability}\label{section:stability}
	We now take a moment to reflect on the significance of our results with regards to the stability of systems with many interacting components, such as complex ecosystems or neural networks. In the case of a complex ecosystem, we take May's approach \cite{may}, and imagine that the ecosystem can be described by a set of differential equations governing the abundances of a large number of interacting species. We suppose we linearise these dynamic equations about a hypothetical fixed point to obtain
	\begin{align}
		\dot x_i = - x_i + \sum_{j}a_{ij} x_j.
	\end{align}
	In this case, $a_{ij}$ are the elements of a Jacobian matrix. Here, we see that if any eigenvalue of the matrix $\underline{\underline{a}}$ has a real part greater than $1$, the equilibrium about which we have linearised is unstable. This allows one to deduce what kinds of interactions between species promote (in)stability. 
	
	In the case of neural networks, we may instead consider the firing-rate model \cite{aljadeff2015transition} where the activation of the $i$th neuron $x_i$ is determined by
	\begin{align}
		\dot x_i = -x_i + \sum_{j}a_{ij} \tanh(x_j),
	\end{align}
	where now $a_{ij}$ describe the connection weights of the neural network. Again, when the leading eigenvalue of $\underline{\underline{a}}$ is greater than 1, the quiescent state at $x_i = 0$ also becomes unstable.
	
	This means that we can interpret the eigenvalue spectrum as follows: if changing the statistics of $a_{ij}$ influences the bulk of the eigenvalue spectrum to expand along the real axis, or it causes the outlier eigenvalue to move further from the origin, the change in the statistics of $a_{ij}$ promotes instability.
	
	We thus are able to draw broad conclusions about the effect of network heterogeneity on stability. The formulae derived in Section \ref{section:circlelaw} in particular (for the boundary of the bulk region and the outlier) are completely independent of the precise network structure and depend only on $s^2$ (the heterogeneity). The approximate results in Section \ref{section:corrections} also allow us to draw general conclusions about the influence of heterogeneity. Although the expressions obtained in Section \ref{section:corrections} are technically only valid for small $s^2$, we verify in Figs. \ref{fig:circlelawdensity}, \ref{fig:bulkedgevsgamma} and \ref{fig:leadingeigenvalue} that the same qualitative behaviour is exhibited for large $s^2$ as for small $s^2$. 
	
	When $\mu$ is small enough that the outlier eigenvalue is not the leading eigenvalue, from Eq.~(\ref{bulkedge}) and Fig. \ref{fig:bulkedgevsgamma}, it is apparent that the effect of network heterogeneity is dependent on $\Gamma$ (the degree to which interactions are symmetric). The quantity plotted in Fig. \ref{fig:bulkedgevsgamma} quantifies the correction to the usual edge of the bulk region of the eigenvalue spectrum. For $\Gamma<1 - \sqrt{2}$ (i.e. very asymmetric `predator-prey' interactions), network degree heterogeneity is a stabilising influence, whereas for $\Gamma > 1-\sqrt{2}$ (i.e. more symmetric interactions), network heterogeneity is destabilising. 
	
	The preceding observation only applies however when $\mu$ is small. If $\mu$ is large enough that the outlier is the leading eigenvalue (see Fig. \ref{fig:leadingeigenvalue}), network heterogeneity acts only to promote instability. This can be seen from Eqs.~(\ref{universaloutlier}) and (\ref{modifiedoutlier}). In short, in most circumstances, network heterogeneity appears to be a destabilising influence, except when the interactions are very asymmetric.
	
	\pagebreak
	
	\section{Outlook}\label{section:conclusion}
	
	In this work, we have studied the eigenvalue spectra of the adjacency matrix of directed complex networks with network heterogeneity that was non-vanishing in the well-connected ($p\to \infty$) limit. In doing so, we were able to make deductions about the effect of network heterogeneity on the stability of a complex system.
	
	Two immediate possibilities for further research using the methods developed here present themselves. Firstly, instead of examining only the limit $p\to \infty$, it would be informative to also involve the possibility of a finite value of $p$ (i.e. a sparse random matrix) \cite{rodgers1988density, semerjian2002sparse}. The eigenvalue spectra of some sparse random matrices with non-trivial network structure have been studied (see e.g. Refs. \cite{nagao2008spectral, rogers2009cavity}), but simple closed-form expressions of the type derived in this work have proved elusive. 
	
	Secondly, the approach developed here could also be extended to examine the spectra of the Laplacian matrices of directed complex networks \cite{biroli1999single}. By examining these spectra, one would gain a deeper insight into the effect that network heterogeneity has on diffusion on complex networks and the phenomenon of localisation \cite{bray1982eigenvalue, baron2021persistent, biroli1999single}. 
	
	\acknowledgements 
	The author acknowledges partial financial support from the Agencia Estatal de Investigaci\'on (AEI, MCI, Spain) and Fondo Europeo de Desarrollo Regional (FEDER, UE), under Project PACSS (RTI2018-093732-B-C21) and the Maria de Maeztu Program for units of Excellence in R\&D, Grant MDM-2017-0711 funded by MCIN/AEI/10.13039/501100011033. 
	
\onecolumngrid
\clearpage
\begin{center}
	\Large{--- Supplemental Material ---}
\end{center}
\setcounter{figure}{0}
\setcounter{equation}{0}
\setcounter{section}{0}
\renewcommand\numberthis{\addtocounter{equation}{1}\tag{\theequation}}
\renewcommand\numberthis{\addtocounter{figure}{1}\tag{\thefigure}}
\renewcommand\numberthis{\addtocounter{section}{1}\tag{\thesection}}
\renewcommand{\thesection}{S\arabic{section}} 		
\renewcommand{\theequation}{S\arabic{equation}}  	
\renewcommand{\thefigure}{S\arabic{figure}}  		
\renewcommand{\thetable}{S\arabic{table}}

\section{Overview}
This document contains additional information about the calculations performed to obtain the results detailed in the main paper. 

To begin with, we perform the calculation discussed in Section III of the main text. Specifically, Section \ref{section:annealedapprox} of this document is dedicated to showing how one can compute the eigenvalue potential, defined in Eq.~(7) of the main text, and to deriving the annealed approximation for the network given in Eq.~(9). Given this annealed approximation, Section \ref{section:saddlepointandresolvent} then shows how the resolvent can be deduced by performing a saddle point approximation (valid for large $N$) of the eigenvalue potential $\Phi(\omega,\omega^\star)$. 

We then move on to deriving the `general results' in Section IV of the main text. Section \ref{section:generalresultsbulk} shows how each of the properties of the bulk of the eigenvalue spectrum (the boundary of the bulk spectrum, the leading eigenvalue of the bulk region and the density of eigenvalues within the bulk region) can be found. These results are valid for a general network degree distribution and are given in Section IV A of the main text. Similarly, Section \ref{section:symmetricmatrices} then shows how one can derive the density of eigenvalues in the case where the random matrix $\underline{\underline{a}}$ is fully symmetric (an undirected network) --- i.e. the results in Section IV B of the main text. Then, Section \ref{section:generalresultsoutlier} derives the general expression for the outlier eigenvalue discussed in Section IV C of the main text

In Section \ref{section:smalls}, we then perform series expansions of the more general results given in Section IV of the main text to obtain the small-$s^2$ (small-network-heterogeneity) modifications to the elliptic, circular and semi-circular laws, as well as the outlier eigenvalues, that were highlighted in Section V of the main text. 

Finally, in Section \ref{section:anys}, we discuss a couple of example degree distributions that were used to produce the figures in the main text, namely the dichotomous and uniform degree distributions. We discuss how the general results in Section IV of the main text can be used to find the properties of the eigenvalue spectrum in these special cases.

\section{Eigenvalue potential and annealed approximation}\label{section:annealedapprox}
In a similar fashion to Refs. \cite{baron2020dispersal, sommers, haake}, we evaluate the eigenvalue potential, defined in Eq.~(7) of the main text, using the replica method \cite{mezard1987}. The replica method exploits the fact that $\ln x = \lim_{n\to 0} (x^n -1)/n$ to evaluate the average of the logarithm in Eq.~(7) by instead calculating the average of an $n$-fold replicated system. 

It has been shown in other works \cite{sommers, haake} (see the Supplemental material in Ref. \cite{baron2022eigenvalues} for a more detailed discussion) that in fact the replicas `decouple', meaning that the logarithm and the ensemble average commute. That is
\begin{align}
	\Phi(\omega, \omega^\star) = - \frac{1}{N} \left\langle\ln \det\left[(\omega^\star \underline{\underline{\id}} - \underline{\underline{a}}^T)(\omega \underline{\underline{\id}} - \underline{\underline{a}}) \right]\right\rangle =- \frac{1}{N} \ln \left\langle\det \left[ (\omega^\star \underline{\underline{\id}} - \underline{\underline{a}}^T)(\omega \underline{\underline{\id}} - \underline{\underline{a}})\right] \right\rangle .
\end{align}
The determinant above can be written as a Gaussian integral (after having performed a Hubbard-Stratonovich transformation \cite{hubbard}) 

\pagebreak

\begin{align}
	\det\left[ (\id \omega^\star - \underline{\underline{a}}^T)(\id \omega- \underline{\underline{a}})\right]^{-1} = &\int \prod_i \frac{d^2z_i d^2y_i}{2 \pi^2} \exp\left[ - \sum_i y_i^\star y_i\right] \nonumber \\
	&\times\exp\left[ i \sum_{ij} z_i^\star (a_{ji} - \omega^\star \delta_{ij})y_j\right] \exp\left[ i \sum_{ij} y_i^\star (a_{ij} - \omega \delta_{ij})z_j\right] .
\end{align}
Performing the ensemble average according to the distribution described in Eqs.~(1) and (2) of the main text, we obtain
\begin{align}
	&\left\langle \exp\left[ i  a_{ij} (z_j^\star  y_i + z_j y_i^\star) + i  a_{ji} (z_i^\star  y_j + z_i y_j^\star)\right]\right\rangle \nonumber \\
	&= 1 + f_{ij} \left[\left\langle \exp\left[ i  a_{ij} (z_j^\star  y_i + z_j y_i^\star) + i  a_{ji} (z_i^\star  y_j + z_i y_j^\star)\right]\right\rangle_\pi-1 \right] \nonumber \\
	&\approx\exp\left[  \frac{k_i k_j}{pN} \left[\left\langle \exp\left[ i  a_{ij} (z_j^\star  y_i + z_j y_i^\star) + i  a_{ji} (z_i^\star  y_j + z_i y_j^\star)\right]\right\rangle_\pi-1 \right] \right]
\end{align}
where we have used the approximation $f_{ij} \approx \frac{k_i k_j}{pN}$ [see Eq.~(3) in the main text], which is valid for $\frac{k_i k_j}{pN} \ll 1$. Now, expanding the exponential and assuming that higher order moments of $a_{ij}$ decay faster then $1/p$, we obtain
\begin{align}
	&\left\langle \exp\left[ i  a_{ij} (z_j^\star  y_i + z_j y_i^\star) + i  a_{ji} (z_i^\star  y_j + z_i y_j^\star)\right]\right\rangle_\pi \nonumber \\
	&\approx 1 + i  \frac{\mu}{p} \left[ (z_j^\star  y_i + z_j y_i^\star) +  (z_i^\star  y_j + z_i y_j^\star) \right] \nonumber \\
	&- \frac{\sigma^2 }{2p} \left[(z_i^{\star} y_j + z_i y_j^{\star})^2 +(z_j^{\star} y_i + z_j y_i^{\star})^2+ 2\Gamma  (z_i^{\star} y_j + z_i y_j^{\star})(z_j^{\star} y_i + z_j y_i^{\star}) \right] \nonumber \\
	&\approx \exp\Bigg[  i  \frac{\mu}{p} \left[ (z_j^\star  y_i + z_j y_i^\star) +  (z_i^\star  y_j + z_i y_j^\star) \right] \nonumber \\
	&- \frac{\sigma^2 }{2p} \left[(z_i^{\star} y_j + z_i y_j^{\star})^2 +(z_j^{\star} y_i + z_j y_i^{\star})^2+ 2\Gamma  (z_i^{\star} y_j + z_i y_j^{\star})(z_j^{\star} y_i + z_j y_i^{\star}) \right]\Bigg].
\end{align}
Finally, we obtain the following approximation for the eigenvalue potential
\begin{align}
	\exp\left[ -N \Phi(\omega) \right] &= \int \prod_{i} \left( \frac{d^2z_i d^2y_i}{2 \pi^2}\right) \exp\left[ - \sum_{i} y^{\star}_i y_i  \right]\nonumber \\
	\times & \exp\left[ -i \sum_{i} \left(z^{\star}_i y_i  \omega^\star + z_i y^{\star}_i  \omega \right) +  i  \frac{\mu}{p} \sum_{ij} \frac{k_i k_j}{pN} (z_i^\star  y_j + z_i y_j^\star) \right]\nonumber \\
	\times & \exp\Bigg[ - \frac{\sigma^2 }{2p} \sum_{ij} \frac{k_i k_j}{pN}\left[(z_i^{\star} y_j + z_i y_j^{\star})^2 + \Gamma  (z_i^{\star} y_j + z_i y_j^{\star})(z_j^{\star} y_i + z_j y_i^{\star}) \right]\Bigg] . \label{averaged}
\end{align}
We note that this is exactly the same expression for $\Phi(\omega,\omega^\star)$ that we would have obtained if we had used the distribution for $a_{ij}$ in Eq.~(9) in the main text from the start. We therefore conclude that the annealed network approximation in Eq.~(9) is valid when $p \gg 1$ and $\frac{k_i k_j}{pN} \ll 1$ for all $i$ and $j$. 

\pagebreak
\section{Finding the resolvent}\label{section:saddlepointandresolvent}
We note from previous works \cite{tao2013outliers, benaych2011eigenvalues, baron2022eigenvalues, orourke} that low-rank perturbations to a random matrix produce outlier eigenvalues, but they do not affect the bulk of the eigenvalue distribution. Noting that the introduction of a non-zero value of $\mu$ is equivalent to a rank-1 perturbation in the annealed approximation, we can set $\mu = 0$ in Eq.~(\ref{averaged}) and continue the calculation to find the bulk eigenvalue density. We return later to the outlier eigenvalue that emerges as a result of setting a non-zero value of $\mu$ in Section \ref{section:generalresultsoutlier}.

\subsection{Introduction of order parameters}
We now imagine that we group each node with all other nodes that share the same degree $k_\alpha$ in a group labelled by the index $\alpha$. In the annealed network approximation the problem therefore reduces to that of finding the eigenvalue spectrum of a random matrix with block-structured statistics (as discussed in the main text). This enables us to follow along the lines of previous works that have studied the eigenvalue spectra of block-structured random matrices (see Ref. \cite{baron2020dispersal} in particular). 

We begin by rewriting Eq.~(\ref{averaged}) as
\begin{align}
	\exp\left[ -N \Phi(\omega) \right] &= \int \prod_{\alpha i} \left( \frac{d^2z^\alpha_i d^2y^\alpha_i}{2 \pi^2}\right) \exp\left[ - \sum_{\alpha i} y^{\alpha\star}_i y^\alpha_i  \right]\nonumber \\
	\times & \exp\left[ -i \sum_{\alpha i} \left(z^{\alpha\star}_i y^\alpha_i  \omega^\star + z^\alpha_i y^{\alpha\star}_i  \omega \right) +  i  \frac{\mu}{p} \sum_{\alpha \beta ij} \frac{k_\alpha k_\beta}{pN} (z_i^{\alpha\star}  y^\beta_j + z^\alpha_i y_j^{\beta\star}) \right]\nonumber \\
	\times & \exp\Bigg[ - \frac{\sigma^2 }{2p} \sum_{\alpha \beta ij} \frac{k_\alpha k_\beta}{pN}\left[(z_i^{\alpha\star} y^\beta_j + z^\alpha_i y_j^{\beta\star})^2 + \Gamma  (z_i^{\alpha\star} y^\beta_j + z^\alpha_i y_j^{\beta\star})(z_j^{\beta\star} y^\alpha_i + z^\beta_j y_i^{\alpha\star}) \right]\Bigg] , \label{averagedalpha}
\end{align}
where the indices $i$ and $j$ now only run over the nodes in the groups $\alpha$ and $\beta$ respectively. We introduce the following `order parameters', defining $x_\alpha = k_\alpha/p$
\begin{align}
	u &= \frac{1}{N}\sum_{\alpha i} x_\alpha z^{\alpha\star}_i z^\alpha_i, \,\,\,\,\, v_\alpha = \frac{1}{N_\alpha}\sum_{i }  y^{\alpha\star}_i y^\alpha_i, \nonumber \\
	w_\alpha &=  \frac{1}{N_\alpha}\sum_i z^{\alpha\star}_i y^\alpha_i, \,\,\,\,\, w_\alpha^\star = \frac{1}{N_\alpha} \sum_{i } y^{\alpha\star}_i z^\alpha_i, \label{orderparameters}
\end{align}
where $N_\alpha$ is the number of nodes with degree $k_\alpha$. We impose these definitions in the integral Eq.~(\ref{averagedalpha}) using Dirac delta functions in their complex exponential representation. We can thus rewrite Eq.~(\ref{averagedalpha}) as 
\begin{align}
	\exp\left[ -N \Phi(\omega) \right] &= \int \mathcal{D}\left[ \cdots \right]\exp\left[N(\Psi + \Theta + \Omega)\right] ,\label{saddlesetup}
\end{align}
where $\mathcal{D}\left[ \cdots \right]$ denotes integration over all of the order parameters and their conjugate (`hatted') variables, and where 
\begin{align}
	\Psi &= i  \hat u u  + i \sum_{\alpha} \gamma_\alpha(\hat v_\alpha  v_\alpha +\hat w_\alpha  w^\star_\alpha  +   \hat w^\star_\alpha  w_\alpha   ), \nonumber \\
	\Theta &= -\sum_\alpha v_\alpha - \sigma^2 u \sum_\alpha \gamma_\alpha x_\alpha v_\alpha - i \sum_\alpha \gamma_\alpha (w_\alpha \omega^\star + w_\alpha^\star \omega) - \frac{\Gamma \sigma^2}{2} \sum_{\alpha \beta} \gamma_\alpha \gamma_\beta x_\alpha x_\beta (w_\alpha w_\beta+ w_\alpha^\star w_\beta^\star) , \nonumber \\
	\Omega &=  \sum_\alpha \gamma_\alpha \ln  \Bigg[\int   \left( \frac{d^2z^\alpha d^2y^\alpha}{2 \pi^2}\right)  \exp\bigg\{ -i  [\hat u x_\alpha z^{\alpha\star} z^\alpha  + \hat v_\alpha   y^{\alpha\star} y^\alpha  \nonumber \\
	& \,\,\,\,\,\,\,\,\,\, + \hat w_\alpha   y^{\alpha\star} z^\alpha + \hat w_\alpha^{\star}  z^{\alpha\star} y^\alpha  ] \bigg\} \Bigg] . \label{argument}
\end{align}
Here, we have defined $\gamma_\alpha = N_\alpha/N$. In the limit $N\to \infty$, this quantity will be given by the degree distribution of the network, i.e. $\gamma_\alpha \to \gamma(k_\alpha)$. 

We note that the integrals over $y_i$ and $z_i$ in Eq.~(\ref{argument}) are uncoupled for different values of $i$ as a result of introducing the order parameters in Eq.~(\ref{orderparameters}). We note also that we have neglected terms involving $N^{-1} \sum_{i} (y^\alpha_i)^2$, $N^{-1} \sum_{i} (z^\alpha_i)^2$, $N^{-1} \sum_{i} z^\alpha_i y^\alpha_i$ and similar terms involving the complex conjugates of $z^\alpha_i$ and $y^\alpha_i$, which do not contribute in the thermodynamic limit (see Refs. \cite{edwardsjones, sommers, haake, baron2022eigenvalues} for further discussion).

Carrying out the integrals over the variables $y_i$ and $z_i$ in the expression for $\Omega$ in Eq.~(\ref{argument}) one obtains
\begin{align}
	\Omega &= -\sum_\alpha \gamma_\alpha \ln\left[\hat w_\alpha  \hat w_\alpha^\star  - x_\alpha \hat u \hat v_\alpha  \right].
\end{align}
\subsection{Saddle point integration and evaluation of the order parameters}
We now suppose that $N \gg 1$, and carry out the integral in Eq.~(\ref{saddlesetup}) in the saddle-point approximation. To do this, we extremise the expression $\Psi + \Theta + \Omega$. Extremising with respect to the conjugate variables $\hat u, \hat v, \hat w_\alpha$ and $\hat w_\alpha^\star$, we find
\begin{align}
	i u &=  - \sum_\alpha\gamma_\alpha x_\alpha  \frac{\hat v_\alpha}{f_\alpha} , \,\,\,\,\, iv_\alpha = -  x_\alpha \frac{\hat u  }{f_\alpha}  , \,\,\,\,\,  i w_\alpha = \frac{\hat w_\alpha}{f_\alpha } , \,\,\,\,\,  i w_\alpha^\star = \frac{\hat w_\alpha^\star}{ f_\alpha} , \nonumber \\
	f_\alpha &= \hat w_\alpha  \hat w_\alpha^\star - x_\alpha \hat u \hat v_\alpha  .\label{diffwrthatted}
\end{align}
We now instead extremise $\Psi + \Theta + \Omega$ in Eq.~(\ref{saddlesetup}) with respect to  $u, v, w_\alpha$ and $w_\alpha^\star$. We find
\begin{align}
	i\hat u &= \sigma^2 v , \,\,\,\,\, i\gamma_\alpha\hat v_\alpha = 1+ \sigma^2 \gamma_\alpha x_\alpha u , \nonumber \\
	i\hat w_\alpha &=  i \omega+ \Gamma\sigma^2 \sum_{\beta} \gamma_\beta x_\alpha x_\beta w^\star_\beta , \,\,\,\,\, i\hat w_\alpha^\star = i\omega^\star + \Gamma\sigma^2 \sum_{\beta} \gamma_\beta x_\alpha x_\beta w_\beta , \label{diffwrtunhatted}
\end{align}
where we introduce $v = \sum_\alpha \gamma_\alpha x_\alpha v_\alpha$.
\subsection{Expression for the resolvent}
In a similar fashion to Ref. \cite{baron2020dispersal}, we observe that the quantities $w_\alpha$ and $w_\alpha^\star$, when evaluated at the saddle point, are related to the resolvent [see Eq.~(8) of the main text]:
\begin{align}
	G(\omega) &= \frac{\partial \Phi(\omega, \omega^\star)}{\partial \omega} = i\sum_\alpha  \gamma_\alpha w^\star_\alpha , \nonumber \\
	G^\star(\omega) &= \frac{\partial \Phi(\omega, \omega^\star)}{\partial \omega^\star} =  i \sum_\alpha \gamma_\alpha w_\alpha . \label{resolventatsaddlepoint}
\end{align}
So, if we can solve Eqs.~(\ref{diffwrthatted}) and (\ref{diffwrtunhatted}) for $i\sum_\alpha  \gamma_\alpha w^\star_\alpha $ as a function of only $\omega$ and $\omega^\star$, we can obtain the eigenvalue density of the bulk region using Eq.~(6) of the main text.

\noindent First, eliminating $\hat u$, one sees that 
\begin{align}
	v = \sigma^2 v \sum_\alpha \gamma_\alpha \frac{x_\alpha^2 }{f_\alpha} . \label{unhatted}
\end{align}

\noindent This equation has two solutions. 
\medskip

\noindent\underline{First solution:}\\
One solution is $v =0$, implying $\hat u = 0$ [see Eq.~(\ref{diffwrtunhatted})], and hence $f_\alpha =  \hat w_\alpha \hat w^\star_\alpha = -1/( w_\alpha w_\alpha^\star)$. We therefore obtain from Eqs.~(\ref{diffwrthatted}) and (\ref{diffwrtunhatted})
\begin{align}
	1 &=    i\omega  w_\alpha^\star+  \Gamma\sigma^2 \sum_\beta \gamma_\beta x_\alpha x_\beta w_\alpha^\star w^\star_\beta , \nonumber \\
	1 &=   i \omega^\star w_\alpha +  \Gamma\sigma^2 \sum_\beta \gamma_\beta x_\alpha x_\beta w_\alpha w_\beta . \label{resolventoutside}
\end{align}
From this, one can solve for the resolvent $G(\omega) = \frac{1}{N}\sum_\alpha  \gamma_\alpha w_\alpha^\star$, which we see is an analytic function of $\omega$. This means that the eigenvalue density vanishes in regions of the complex plane for which $v=0$ is the only valid solution [as a result of Eq.~(6) of the main text].

\medskip

\noindent\underline{Second solution:} \\
The other solution to the first of Eq.~(\ref{unhatted}) is 
\begin{align}
	\sum_\alpha \gamma_\alpha \frac{ x_\alpha^2}{f_\alpha} = \frac{1}{\sigma^2} . \label{conditionforinside}
\end{align}
Now, we see from the expression for $iu$ and $i \hat v_\alpha$ in Eqs.~(\ref{diffwrthatted}) and (\ref{diffwrtunhatted}) that when the second Eq.~(\ref{conditionforinside}) is satisfied, $u\to \infty$. This means that $i\hat v_\alpha \to \sigma^2 x_\alpha u$ and hence
\begin{align}
	\hat v_\alpha \hat u \to  x_\alpha^2 g(\omega, \omega^\star),  
\end{align}
where $g(\omega, \omega^\star)$ is an arbitrary function to be found that is independent of $\alpha$. Hence, we obtain the following simultaneous equations, which enable us to find the resolvent $G(\omega, \omega^\star) = \sum_{\alpha} \gamma_\alpha (i w_\alpha^\star)$
\begin{align}
	\frac{1}{\sigma^2} &= \sum_\alpha \gamma_\alpha \frac{x_\alpha^2}{f_\alpha} , \nonumber \\
	- w_\alpha w_\alpha^\star &= 1/f_\alpha + x_\alpha^2 \frac{g(\omega,\omega^\star)}{f_\alpha^2}, \nonumber \\
	iw_\alpha  &=   \frac{\omega}{f_\alpha}-  \Gamma\sigma^2\frac{x_\alpha}{f_\alpha} \sum_{\beta}  iw^\star_\beta \gamma_\beta  x_\beta  ,\nonumber \\ 
	iw_\alpha^\star &= \frac{\omega^\star}{f_\alpha}-  \Gamma\sigma^2 \frac{x_\alpha}{f_\alpha} \sum_{\beta} iw_\beta \gamma_\beta x_\beta . \label{simfordens}
\end{align}
In principle, one can solve these along with Eq.~(\ref{conditionforinside}) to find $g(\omega, \omega^\star)$, $iw_\alpha$ and $iw_\alpha^\star$ as functions of $\omega$ and $\omega^\star$. In this case, the resolvent is no longer necessarily an analytic function of $\omega$. Therefore, in the region of the complex plane where Eq.~(\ref{conditionforinside}) is satisfied, the eigenvalue density is non-zero.

Noting Eq.~(\ref{resolventatsaddlepoint}) and Eq.~(6) from the main text, one then obtains the eigenvalue density via
\begin{align}
	\rho(\omega) = \frac{1}{\pi}\mathrm{Re}\left[\sum_\alpha \gamma_\alpha \frac{\partial iw_\alpha^\star}{\partial \omega^\star} \right] . \label{densityfromres}
\end{align}

\section{General results for the bulk region}\label{section:generalresultsbulk}
We have discussed two solutions to Eq.~(\ref{unhatted}): one that corresponds to the region of the complex plane in which the bulk of the eigenvalue spectrum of $\underline{\underline{a}}$ resides and one that corresponds to the region of the complex plane where there are no eigenvalues. Noting these two solutions [given in Eqs.~(\ref{resolventoutside}) and (\ref{simfordens})] along with Eq.~(\ref{densityfromres}), we are now in a position to derive the results given in Section IV A of the main text.

\subsection{Replacing the index $\alpha$ with $k$}
So as to make clear the effective block structure of the random matrix that we were considering, we introduced the index $\alpha$, which indexed nodes with the same degree $k_\alpha$. We now use the fact that $\gamma_\alpha = \gamma(k_\alpha)$ in the large $N$ limit and drop the index $\alpha$ to obtain the expressions in Section IV A of the main text. For example, the expressions in Eqs.~(\ref{simfordens}) become
\begin{align}
	\frac{1}{\sigma^2} &= \sum_k \gamma(k) \frac{(k/p)^2}{f_k}, \nonumber \\
	-w_k w_k^\star &= 1/f_k + (k/p)^2 \frac{g(\omega,\omega^\star)}{f_k^2}, \nonumber \\
	iw_k  &=   \frac{\omega}{f_k}-  \Gamma\sigma^2\frac{(k/p)}{f_k} \sum_{k'} \gamma(k')  (k'/p) (iw^\star_{k'}) ,\nonumber \\ 
	iw_k^\star &= \frac{\omega^\star}{f_k}-  \Gamma\sigma^2 \frac{(k/p)}{f_k} \sum_{k'} \gamma(k') (k'/p) (iw_{k'}) . 
\end{align}

\subsection{Boundary of the bulk region}
We first derive the results in Section IV A 1 of the main text for the boundary of the bulk region. Because the two solutions to Eq.~(\ref{unhatted}) correspond to the region inside the bulk of the eigenvalue spectrum and the outside, the boundary of the bulk region is given by the set of points $\omega = \omega_x + i \omega_y$ that simultaneously satisfy both Eqs.~(\ref{resolventoutside}) and (\ref{simfordens}). The simultaneous solution of these equations yields
\begin{align}
	\frac{1}{\sigma^2} &= \sum_k \gamma(k) (k/p)^2 \vert iw_k^\star \vert^2, \nonumber \\
	i w_k^\star &= \frac{1}{i \omega + \Gamma \sigma^2 (k/p) \sum_{k'} \gamma(k') (k'/p) [i w_{k'}^\star]} .
\end{align}
Let $\sum_k \gamma(k) (k/p) (i w_k^\star) = A_x + i A_y$, where both $A_x$ and $A_y$ are real. Then we obtain
\begin{align}
	\frac{1}{\sigma^2} &= \sum_k \gamma(k) \frac{(k/p)^2 }{[\omega_x - \Gamma \sigma^2 (k/p) A_x]^2+ [\omega_y - \Gamma \sigma^2 (k/p) A_y]^2}, \nonumber \\
	A_x &= \sum_k \gamma(k) \frac{(k/p) \omega_x - \Gamma \sigma^2 (k/p)^2 A_x }{[\omega_x - \Gamma \sigma^2 (k/p) A_x]^2+ [\omega_y - \Gamma \sigma^2 (k/p) A_y]^2}, \nonumber \\
	A_y &= \sum_k \gamma(k) \frac{-(k/p) \omega_y + \Gamma \sigma^2 (k/p)^2 A_y }{[\omega_x - \Gamma \sigma^2 (k/p) A_x]^2+ [\omega_y - \Gamma \sigma^2 (k/p) A_y]^2}. \label{bulkboundary}
\end{align}
Now defining
\begin{align}
	h = \sum_k \gamma(k) \frac{k/p}{[\omega_x - \Gamma \sigma^2 (k/p) A_x]^2+ [\omega_y - \Gamma \sigma^2 (k/p) A_y]^2}, 
\end{align}
and noting that $A_x = \omega_x h/(1 + \Gamma)$ and $A_y = -\omega_yh/(1-\Gamma)$, one arrives at the expressions in Eq.~(10) of the main text.

\subsection{Leading eigenvalue of the bulk region}
The leading eigenvalue of the bulk region can be obtained by finding the point on the boundary of the bulk region with $\omega_y = 0$. Making this substitution in Eqs.~(\ref{bulkboundary}), one readily obtains Eqs.~(13) in the main text.

\subsection{Eigenvalue density in the bulk region}
We now derive the results in Section IV A 3 of the main text for the eigenvalue value density inside the bulk region. Defining $m = \sum_{k} \gamma(k)\frac{(k/p)}{f_k}$, one finds after some rearrangement of the last two of Eqs.~(\ref{simfordens}) that
\begin{align}
	iw_k^\star &= \frac{\omega^\star}{f_k} - \Gamma \sigma^2 \frac{(k/p)}{f_k} \frac{(\omega - \Gamma \omega^\star)}{1-\Gamma^2} m. \label{wexpression}
\end{align}
Thus, substituting this into the second of Eqs.~(\ref{simfordens}), one obtains
\begin{align}
	f_k = \left\vert \omega^\star - \Gamma \sigma^2 (k/p)\frac{(\omega - \Gamma \omega^\star)}{1-\Gamma^2} m  \right\vert^2 - (k/p)^2 g(\omega, \omega^\star) . \label{fexpression}
\end{align}
Using Eq.~(\ref{wexpression}) in combination with Eq.~(\ref{fexpression}) and Eq.~(\ref{densityfromres}), one obtains Eq.~(14) in the main text. Also substituting Eqs.~(\ref{wexpression}) and (\ref{fexpression}) the first of Eqs.~(\ref{simfordens}) and the definition of $m$ above, one obtains Eqs.~(15) in the main text. 

\section{General results for symmetric matrices}\label{section:symmetricmatrices}
In the case of a symmetric matrix, all the eigenvalues lie on the real axis. For this reason, it is no longer useful to consider the eigenvalue density, as defined in Eq.~(4) of the main text. This definition has the normalisation $\int d^2 \omega \rho(\omega) = 1$, where the integral is taken over the whole complex plane. Instead, we now define the real eigenvalue density to be normalised such that $\int d\omega_x \rho_x(\omega_x) = 1$, where we still have
\begin{align}
	\rho_x(\omega_x) = \left\langle \frac{1}{N} \sum_i \delta(\omega - \lambda_i) \right\rangle,
\end{align}
but now the delta functions are taken to have only a real argument. In this case, one instead obtains the eigenvalue density from the trace of the resolvent matrix via \cite{taobook, edwardsjones}
\begin{align}
	\rho_x(\omega_x) = -\frac{1}{\pi} \lim_{\epsilon \to 0} \mathrm{Im}\left[ G(\omega_x+i\epsilon)\right]. 
\end{align}
The definition of $G$ remains the same as in Eq.~(5) of the main text. However, because we only consider real values of $\omega = \omega_x$, the resolvent must be the analytic \cite{JANIK1997603}. When this is the case, $i w_k^\star$ satisfies Eqs.~(\ref{resolventoutside}). Solving Eqs.~(\ref{resolventoutside}) for $iw_k^\star$ and substituting this into Eq.~(\ref{resolventatsaddlepoint}), one thus obtains Eqs.~(16) and (17) in the main text.

\section{General results for the outlier eigenvalues}\label{section:generalresultsoutlier}
So far, we have discussed how one can deduce the properties of the bulk region of the eigenvalue spectrum of $\underline{\underline{a}}$, to which most of the eigenvalues of confined. We did this by setting $\mu = 0$, which only has the effect of removing the outlier eigenvalue \cite{tao2013outliers, baron2020dispersal, baron2022eigenvalues} without affecting the bulk region. We now reintroduce a non-zero value of $\mu$ and deduce the location of the outlier eigenvalue in the complex plane.

We begin with Eq.~(19) of the main text, which is simply the definition of an eigenvalue of the matrix $\underline{\underline{a}}$,
\begin{align}
	\det \left[ \lambda_{\mathrm{outlier}} \underline{\underline{\id}} - \underline{\underline{z}} - p^{-1} \underline{\underline{\mu}}\right] = 0, \label{detoutlier}
\end{align}
where we define $\underline{\underline{z}} = \underline{\underline{a}} - p^{-1}\underline{\underline{\mu}}$, with $\underline{\underline{\mu}} = p \left\langle \underline{\underline{a}} \right\rangle$. 

The elements of the matrix $\underline{\underline{\mu}}$, as was shown in Section \ref{section:annealedapprox} [see also Eq.~(9) of the main text], can be replaced by $\mu \frac{k_i k_j}{N}$. This is a rank-1, block-structured matrix. Let us group nodes with the same degree and introduce a block index (superscript) so that $\mu_{ij}^{kl} = \frac{k l}{ N}$ is the element in $i$th row, the $j$th column of the block in the $k$th row of blocks and the $l$th row of blocks. 

As was mentioned in the main text, we see that we can rewrite Eq.~(\ref{detoutlier}) as
\begin{align}
	\det\left[ \underline{\underline{\id}} - p^{-1} \underline{\underline{G_0}} \underline{\underline{\mu}}\right] &= 0, \label{zeroresoutlier}
\end{align}
where we write $\underline{\underline{G_0}} =  \left[\lambda_{\mathrm{outlier}} \underline{\underline{\id}} - \underline{\underline{z}}\right]^{-1}$.

We follow the reasoning in Ref. \cite{baron2020dispersal}, which also deals with block structured matrices (see Section S3 of the Supplemental Material of this reference in particular). There, it is demonstrated for general block-structured matrices that the resolvent matrix $\underline{\underline{G_0}}$ is diagonal and has elements $G_{ij}^{kl} = \delta_{ij}\delta_{kl} G_k$, where $G_k = i w_k^\star$ is the contribution to the resolvent corresponding to the $k$th block. The resolvent is evaluated outside the bulk region, since we are dealing with an outlier eigenvalue, so $i w_k^\star$ is given by the expression in Eq.~(\ref{resolventoutside}).

Now, we use Sylvester's determinant identity 
\begin{align}
	\det\left[\id_m + \underline{\underline{A}} \underline{\underline{B}} \right] = \det\left[\id_k + \underline{\underline{B}} \underline{\underline{A}} \right], 
\end{align}
which is valid for combinations of $m \times k$ matrices $\underline{\underline{A}} $ and $k \times m$ matrices $\underline{\underline{B}} $. Noting that $\underline{\underline{\mu}}$ can be written as a product of two vectors of dimension $N \times 1$ and $1 \times N$, i.e. $\underline{\underline{\mu}} = \mu (pN)^{-1}\underline{v} \underline{v}^T$ with $v_i^k = k$, one finds from Eq.~(\ref{zeroresoutlier})
\begin{align}
	\det\left[ \underline{\underline{\id}}_N - p^{-1} \underline{\underline{G_0}} \underline{\underline{\mu}}\right] = \det\left[ \underline{\underline{\id}}_1 - \mu (p^2N)^{-1}\underline{v}^T \underline{\underline{G_0}} \underline{v}\right] &= 0, \nonumber \\
	\Rightarrow 1 - \mu\sum_k \gamma(k)(k/p)^2 (i w_k^\star) &= 0,
\end{align}
where we have used $N_k/N \to \gamma(k)$ when $N\to \infty$, where $N_k$ is the number of nodes with degree $k$. Using Eqs.~(\ref{resolventoutside}), one thus obtains Eqs.~(21) of the main text.

\section{Corrections to known results for non-zero network heterogeneity}\label{section:smalls}
In Sections \ref{section:generalresultsbulk}, \ref{section:symmetricmatrices} and \ref{section:generalresultsoutlier}, we derived general expressions for the bulk region and the outlier eigenvalue, which are valid for an arbitrary network degree distribution $\gamma(k)$. These expressions, while useful, are not always easy to evaluate and they do not offer us an intuitive understanding of the effects of a non-trivial complex network structure on the eigenvalue spectrum. 

In this section, we derive the approximations to the eigenvalue spectrum discussed in Section V of the main text. These approximations are valid for small values of the network heterogeneity $s^2$, which is defined as
\begin{align}
	s^2 = \sum_k \gamma(k) (k-p)^2/p^2 .
\end{align}
\subsection{Bulk region}
\subsubsection{$\Gamma = 0$: Universal circular law and bulk density }
In this section, we derive Eqs.~(26)--(29) in the main text, which gives the eigenvalue density in the case $\Gamma = 0$. Inside the bulk region of the eigenvalue spectrum, the boundary of which is given by Eq.~(24) of the main text, the resolvent is given by Eqs.~(14) and (15) in the main text. In the special case $\Gamma = 0$, one obtains 
\begin{align}
	G(\omega, \omega^\star) &= \sum_k \gamma(k)\frac{\omega^\star}{\vert \omega\vert^2 - (k/p)^2 g(\vert \omega\vert)}, \label{resolventfromg}
\end{align}
where the function $g(\vert \omega\vert)$ is obtained by solving
\begin{align}
	\frac{1}{\sigma^2} &= \sum_k \gamma(k) \frac{(k/p)^2}{\vert \omega\vert^2 - (k/p)^2 g(\vert \omega\vert)} . \label{gsol}
\end{align}
Noting that $\rho(\vert\omega\vert) = \frac{1}{\pi} \mathrm{Re} \left\{\frac{\partial}{\partial \omega^\star} G(\omega, \omega^\star) \right\}$ [see Eq.~(6) of the main text], we obtain for the eigenvalue density
\begin{align}
	\rho(\vert \omega\vert) = \frac{1}{\pi} \mathrm{Re} \left\{\sum_k \gamma(k)\frac{1}{\vert \omega\vert^2 - (k/p)^2 g(\vert \omega\vert)} - \sum_k \gamma(k) \left[\omega -  (k/p)^2 \frac{\partial g}{\partial \omega^\star}\right]\frac{\omega^\star}{[\vert \omega\vert^2 - (k/p)^2 g(\vert \omega\vert)]^2} \right\} . \label{diffedrho}
\end{align}
Differentiating Eq.~(\ref{gsol}), we also obtain
\begin{align}
	\frac{\partial g}{\partial \omega^\star} \sum_k \gamma(k) \frac{(k/p)^4}{[\vert \omega\vert^2 - (k/p)^2g]^2} = \omega \sum_k \gamma(k) \frac{(k/p)^2}{[\vert \omega\vert^2 - (k/p)^2g]^2} . \label{diffedg}
\end{align}
Eliminating $\partial g/\partial \omega^\star$ from Eq.~(\ref{diffedrho}) using Eq.~(\ref{diffedg}), we then arrive Eq.~(26) in the main text.

Now we turn our attention to the small-$s^2$ expansion in Eq.~(29) of the main text. Noting that $g(\vert\omega\vert)$ fully determines the eigenvalue density, we merely need to find an approximation for $g(\vert\omega\vert)$ up to first order in $s^2$ and insert this into Eq.~(\ref{resolventfromg}) to obtain the eigenvalue density. 

We suppose that we can approximate $g \approx g_0 + s^2 g_1$. We also make the substitution $k = p (1 + \Delta_k)$, so that $\sum_k \gamma(k) (k-p)^2/p^2 = \sum_k \gamma(k) \Delta_k^2 = s^2$. Expanding the summand of Eq.~(\ref{gsol}) as a series in $\Delta_k$, carrying out the sums over $k$ and equating terms of the same order in $s^2$ on either side, we then have
\begin{align}
	\frac{1}{\vert \omega\vert^2 - g_0 } &= \frac{1}{\sigma^2} , \nonumber \\
	\vert \omega\vert^4 + 3 g_0 \vert \omega\vert^2 + g_1 \vert \omega\vert^2 - g_0 g_1 &= 0 .
\end{align}
One can solve these equations simultaneously to obtain
\begin{align}
	g_0 &= \vert \omega\vert^2 - \sigma^2, \nonumber \\
	g_1 &= \frac{\vert \omega\vert^2 }{\sigma^2} (3 \sigma^2 - 4 \vert \omega\vert^2). \label{g0g1}
\end{align}
Now, expanding the right-hand side of Eq.~(\ref{resolventfromg}) in a similar way, we obtain
\begin{align}
	G(\omega, \omega^\star) &\approx \frac{\omega^\star}{\vert \omega\vert^2 - g_0 } + s^2 \omega^\star \frac{g_0 \vert \omega\vert^2+  3 g_0^2 + g_1 \vert \omega\vert^2 - g_0 g_1 }{(\vert \omega\vert^2 - g_0 )^3}. \label{gexpand}
\end{align}
Substituting the expressions for $g_0$ and $g_1$ in Eq.~(\ref{g0g1}) into Eq.~(\ref{gexpand}), we finally obtain
\begin{align}
	G(\omega, \omega^\star) = \frac{\omega^\star}{\sigma^2}\left[1 + s^2 \left(3  - 4 \frac{\vert\omega\vert^2}{\sigma^2}\right)\right],
\end{align}
from which one recovers Eq.~(29) of the main text using Eq.~(6), after dividing through by an appropriate normalising factor [which doesn't affect the approximation to order $O(s^2)$].
\subsubsection{$\Gamma \neq 0$ and $\Gamma \neq 1$: Modified elliptic law for small $s^2$}
Now, we demonstrate how Eq.~(30) of the main text [the small-$s^2$ approximation for the boundary of the bulk of the eigenvalue spectrum for arbitrary $\Gamma$] can be derived from Eqs.~(10). Our aim is to find $\omega_y$ as a function of $\omega_x$ to first order in $s^2$ in such a way that the leading eigenvalue of the bulk region is also correctly predicted to first order in $s^2$ [this is given in Eq.~(31) of the main text]. To this end, we perform a similar expansion as in the previous subsection. We do this by expanding the quantities $h(\omega_x)$ and $\omega_y(\omega_x)$ to first order in $s^2$. 

Expanding Eqs.~(10) of the main text, we have up to first order in $s^2$
\begin{align}
	\frac{1}{\sigma^2} &\approx \frac{1}{\omega_x^2 \left[ 1 - \frac{\Gamma \sigma^2}{(1+ \Gamma) } (h_0 + s^2 h_1)\right]^2 + \left[\omega^{(0)}_y + \frac{s^2}{2} \omega^{(1)}_y \right]^2  \left[ 1 + \frac{\Gamma \sigma^2}{(1- \Gamma) }(h_0 + s^2 h_1)\right]^2} + s^2 \frac{f_\sigma}{\sigma^2} , \nonumber \\
	h &\approx \frac{1}{\omega_x^2 \left[ 1 - \frac{\Gamma \sigma^2}{(1+ \Gamma)}  (h_0 + s^2 h_1)\right]^2 + \left[\omega^{(0)}_y + \frac{s^2}{2} \omega^{(1)}_y \right]^2  \left[ 1 + \frac{\Gamma \sigma^2}{(1- \Gamma)} (h_0 + s^2 h_1)\right]^2} + s^2 \frac{f_h}{\sigma^2}.  \label{fdef}
\end{align}
We note that we have included contributions of the order $s^2$ to the denominator of the leading order terms so as to correctly preserve the critical value of $\omega_y^2$ at which $\omega_y^2= 0$, in a similar way to the procedure in Refs. \cite{kim1985density, rodgers1988density}. Eliminating $h$ in Eqs.~(\ref{fdef}), we then find (writing simply $\left[\omega^{(0)}_y + \frac{s^2}{2} \omega^{(1)}_y \right]^2 \approx \omega_y^2$, understanding that $\omega_y^2$ is approximate to first order in $s^2$)
\begin{align}
	\omega_x^2 \left[ \frac{1 }{(1+ \Gamma) }   - s^2 \frac{\Gamma }{(1+ \Gamma) }  (f_h - f_\sigma)\right]^2 + \omega_y^2 \left[ \frac{1}{(1- \Gamma) } + s^2\frac{\Gamma }{(1- \Gamma) }  (f_h - f_\sigma)\right]^2 \approx \sigma^2 (1 + s^2 f_\sigma).\label{frearrange}
\end{align}
This hints at the form of the solution. Clearly, we will end up with some sort of modified ellipse. Now, by performing the expansion of Eqs.~(10) in the main text, as in the previous section, and comparing coefficients of $s^2$ with Eqs.~(\ref{fdef}), one obtains 
\begin{align}
	f_h &= f_\sigma + \frac{2\Gamma}{\sigma^2} \left[ \frac{\omega_y^2}{(1-\Gamma)^2} - \frac{\omega_x^2}{(1+\Gamma)^2} \right] - 1, \nonumber \\
	f_\sigma &= 1 - \frac{4 \Gamma}{\sigma^2} \left[\frac{\omega_y^2}{(1-\Gamma)^2}- \frac{\omega_x^2}{(1+\Gamma)^2} \right] - \frac{\Gamma^2}{\sigma^2} \left[\frac{\omega_y^2}{(1-\Gamma)^2}+ \frac{\omega_x^2}{(1+\Gamma)^2} \right] \nonumber \\
	&+ \frac{4 \Gamma^2}{\sigma^4} \left[\frac{\omega_y^2}{(1-\Gamma)^2}- \frac{\omega_x^2}{(1+\Gamma)^2} \right]^2, \label{fexpressions}
\end{align}
where we have used
\begin{align}
	\frac{1}{\sigma^2} &\approx h_0 \approx \frac{1}{\omega_x^2 \left[ 1 - \frac{\Gamma \sigma^2}{(1+ \Gamma) } h \right]^2 + \omega_y^2  \left[ 1 + \frac{\Gamma \sigma^2}{(1- \Gamma) }h\right]^2} 
\end{align}
in the coefficient of $s^2$ (this does affect our approximation to first-order in $s^2$). 

Expanding Eq.~(\ref{frearrange}) to first order in $s^2$ using Eqs.~(\ref{fexpressions}), quartic terms in $\omega_x$ and $\omega_y$ cancel and one obtains
\begin{align}
	\frac{\omega_x^2}{(1 + \Gamma)^2} \left[ 1 - s^2 (1 + 2 \Gamma - \Gamma^2) \right] + \frac{\omega_y^2}{(1 - \Gamma)^2} \left[ 1 - s^2 (1 - 2 \Gamma - \Gamma^2) \right] = \sigma^2. \label{modellipticlaw}
\end{align}
Finally, noting that $1/(1 + x)^2 \approx 1 + 2 x$ for small $x$, we arrive at Eq.~(30) in the main text.

We note that if we set $\omega_y = 0$ in Eq.~(\ref{modellipticlaw}), we obtain $\omega_x = \sqrt{(1 + \Gamma) +s^2 (1 + 3\Gamma + \Gamma^2 - \Gamma^3)} \approx (1 + \Gamma) +s^2 (1 + 3\Gamma + \Gamma^2 - \Gamma^3)/2$. We can compare this result with the expansion of Eq.~(13) of the main text. Letting $\lambda_\mathrm{edge} \approx \lambda_0 + s^2 \lambda_1$ and $A = A_0 + s^2 A_1$, we obtain by comparing coefficients of $s^2$ in this expansion 
\begin{align}
	\frac{1}{\sigma^2} & = \frac{1}{(\lambda_0 - A_0 \Gamma \sigma^2)^2}, \nonumber \\
	0 &= \lambda_0^2 + A_0 \Gamma \sigma^2 \lambda_0 - 2 \lambda_0 \lambda_1 + 2 A_1 \Gamma \sigma^2 \lambda_0 + 2 A_0 \Gamma \sigma^2 \lambda_1 - 2 A_0 A_1 \Gamma^2 \sigma^4, \nonumber \\
	A_0 &= \frac{1}{\lambda_0 - A_0 \Gamma \sigma^2} , \nonumber \\
	A_1 &= \frac{A_0 \Gamma \sigma^2 \lambda _0 - \lambda_0 \lambda_1 + A_1 \Gamma \sigma^2 \lambda_0 + A_0 \Gamma \sigma^2 \lambda_1 - A_0 A_1 \Gamma^2 \sigma^4}{(\lambda_0 - A_0 \Gamma \sigma^2)^3}.
\end{align}
Solving these simultaneously, we obtain
\begin{align}
	A_0 &= \frac{1}{\sigma}, \nonumber \\
	\lambda_0 &= (1 + \Gamma) \sigma, \nonumber \\
	A_1&= -\frac{(1 +\Gamma)^2}{2 \sigma}, \nonumber \\
	\lambda_1 &= \frac{\sigma}{2} (1 + 3 \Gamma + \Gamma^2 - \Gamma^3), 
\end{align}
which agrees with the expression in Eq.~(31) of the main text. This means that the expansion we performed to obtain the modified elliptic law in Eq.~(30) of the main text correctly preserved the point at which $\omega_y \to 0$ to first order in $s^2$, as desired. 
\subsubsection{$\Gamma = 1$: Modified semi-circular law}
We now derive the modified semi-circular law in Eq.~(33) of the main text. This is a first-order (in $s^2$) approximation to the eigenvalue density along the real axis in the case where $\Gamma = 1$, where we also preserve the point $\omega_x = \omega_c$ at which the eigenvalue density goes to zero to first order in $s^2$.

We begin with Eqs.~(17) in the main text and expand in a similar way to the previous subsection to obtain
\begin{align}
	A &\approx \frac{1}{\omega_x - \sigma^2 A} + s^2 \frac{A  \sigma^2 \omega_x}{(\omega_x - \sigma^2 A)^3} , \nonumber \\
	G &\approx \frac{1}{\omega_x - \sigma^2 A} + s^2 \frac{A^2  \sigma^4}{(\omega_x - \sigma^2 A)^3} . \label{semicircleexp1}
\end{align}
Now the aim is to obtain an expression for $A$ that is accurate to first order in $s^2$ and that preserves the square root singularity of the eigenvalue density at the edge of the bulk of the eigenvalue spectrum. The procedure we use is similar to Ref. \cite{kim1985density}.

We begin by noting that the zeroth-order approximation for $A$ satisfies 
\begin{align}
	\sigma^2 A_0^2 - \omega_x A_0 +1=0 .
\end{align}
Using this, we can rewrite the coefficient of $s^2$ in the first of Eqs.~(\ref{semicircleexp1}) to obtain
\begin{align}
	\sigma^2 A^2 - \omega_x A +1 - s^2 \omega_x (1 - \omega_x A) A \approx 0.
\end{align}
We can thus solve this quadratic expression for $A$ and find
\begin{align}
	A = \frac{1}{2( \sigma^2 + s^2 \omega_x^2)} \left[ \omega_x (1 + s^2) - \sqrt{\omega_x^2 (1 - 2 s^2 + s^4) - 4 \sigma^2} \right] . \label{asol}
\end{align}
Similarly, we find for the trace of the resolvent
\begin{align}
	G \approx A - s^2 \sigma^2 A^3 . \label{gexpression}
\end{align}
Noting Eq.~(16) in the main text , we see that one only obtains a non-zero eigenvalue density when the argument of the radical in Eq.~(\ref{asol}) is negative. We thus see that the critical value of $\omega^2$ at which the eigenvalue density switches from non-zero to zero is (to first order in $s^2$)
\begin{align}
	\omega^2_c \approx 4 \sigma^2 (1 + 2s^2) .
\end{align}
Since we only wish to preserve this critical value to leading order in $s^2$, we can ignore the term proportional to $s^4$ in Eq.~(\ref{asol}). We can thus rewrite $A$ to leading order in $s^2$ as
\begin{align}
	A = \frac{2}{\omega^2_c} \left[ 1 - 2 s^2 \left(\frac{2\omega^2}{\omega^2_c}-1\right) \right]\left[ \omega (1 + s^2) - (1 - s^2 )\sqrt{\omega^2  - \omega^2_c} \right] . 
\end{align}
We use this expression to find an approximation for $A^3$ that is valid to first order in $s^2$ and that preserves the singularity at $\omega_c^2$. We find
\begin{align}
	s^2 \sigma^2 A^3 \approx  s^2 \frac{2}{\omega_c^4} \left[ \omega^3 + 3 \omega (\omega^2 - \omega_c^2) -(4 \omega^2 - \omega_c^2) \sqrt{\omega^2 - \omega_c^2}\right] .
\end{align}
Finally, substituting this expression into Eq.~(\ref{gexpression}) and using Eq.~(16) in the main text, we arrive at the modified semi-circular law in Eq.~(33).
\subsection{Outlier eigenvalues: approximate expression for small $s^2$ and general $\Gamma$}
In this subsection, we begin with Eqs.~(21) in the main text and derive the approximate expression for the outlier eigenvalue in Eq.~(32) of the main text, which is accurate to first order in $s^2$. In a similar spirit to the previous section, we imagine that we can write $A \approx A_0 + s^2 A_1$ and $\lambda_\mathrm{outlier} \approx \lambda_0 + s^2 \lambda_1$. We then expand Eqs.~(21) and equate terms with the same power of $s^2$ to obtain
\begin{align}
	A_0 &= \frac{1}{\lambda_0 - A_0 \Gamma \sigma^2}, \nonumber \\
	A_1 &= \frac{ A_0 \Gamma \lambda_0 \sigma^2 - \lambda_0 \lambda_1 + A_1 \Gamma \lambda_0 \sigma^2 + A_0 \Gamma \lambda_1 \sigma^2 - A_0 A_1 \Gamma^2 \sigma^4 }{\left[\lambda_0 - A_0 \Gamma \sigma^2 \right]^3}, \nonumber \\
	\frac{1}{\mu}&= \frac{1}{\lambda_0 - A_0 \Gamma \sigma^2}, \nonumber \\
	0 &= \frac{\lambda_0^2 - \lambda_0 \lambda_1 + A_1 \Gamma \lambda_0 \sigma^2 + A_0 \Gamma \lambda_1 \sigma^2 - A_0 A_1 \Gamma ^2 \sigma^4}{\left[\lambda_0 - A_0 \Gamma \sigma^2 \right]^3} .  \label{outliersimulataneous}
\end{align}
From the first and third of these equations, one thus finds $A_0 = 1/\mu$ and $\lambda_0 = \mu + \Gamma \sigma^2/\mu$. Substituting these expressions into the second and fourth of Eqs.~(\ref{outliersimulataneous}), one finds
\begin{align}
	A_1 &= \frac{1}{\mu} + s^2 \frac{\Gamma \sigma^2}{\mu^3}, \nonumber \\
	\lambda_1 &= \mu + \frac{\Gamma \sigma^2}{\mu}.
\end{align}
Combining the expressions for $\lambda_0$ and $\lambda_1$ above, we arrive at Eq.~(32) of the main text.
\section{Some examples that are valid for any value of $s^2$}\label{section:anys}
\subsection{Dichotomous degree distribution}
To produce the solid lines in Figs. 1 and 2a, a dichotomous degree distribution was used with 
\begin{align}
	\gamma(k) = \frac{1}{2}\left(\delta_{k, k_1} + \delta_{k, k_2} \right) .
\end{align}
In the case of Fig. 2a, all that one is required to calculate from this distribution is the mean degree and heterogeneity, which are given by respectively
\begin{align}
	p &= \frac{1}{2}(k_1 + k_2), \nonumber \\
	s^2 &= \frac{1}{2 p^2}\left[(k_1 - p)^2 + (k_2-p)^2\right] .
\end{align}
In the case of Fig. 1, one must solve Eqs.~(10) of the main text. That is, for each value of $\omega_x$, one must first solve the following simultaneous equations (this is best done numerically) for $h$ and $\omega_y$
\begin{align}
	h = \frac{1}{2p} \left[ \frac{k_1}{\omega_x^2[1 - \frac{\Gamma \sigma^2 k_1}{(1+\Gamma)p} h]^2 + \omega_y^2[1 + \frac{\Gamma \sigma^2 k_1}{(1-\Gamma)p} h]^2 }+  \frac{k_2}{\omega_x^2[1 - \frac{\Gamma \sigma^2 k_2}{(1+\Gamma)p} h]^2 + \omega_y^2[1 + \frac{\Gamma \sigma^2 k_2}{(1-\Gamma)p} h]^2} \right], \nonumber \\
	\frac{1}{\sigma^2} = \frac{1}{2p^2} \left[ \frac{k_1^2}{\omega_x^2[1 - \frac{\Gamma \sigma^2 k_1}{(1+\Gamma)p} h]^2 + \omega_y^2[1 + \frac{\Gamma \sigma^2 k_1}{(1-\Gamma)p} h]^2 }+  \frac{k_2^2}{\omega_x^2[1 - \frac{\Gamma \sigma^2 k_2}{(1+\Gamma)p} h]^2 + \omega_y^2[1 + \frac{\Gamma \sigma^2 k_2}{(1-\Gamma)p} h]^2} \right].
\end{align}

\subsection{Uniform degree distribution}
The dichotomous distribution discussed above is fairly straightforward to implement. The uniform distribution used in Figs. 2b, 3, 4, 5 and 6 requires some additional manipulation. 

The degree distribution is given by
\begin{align}
	\gamma(k) = \frac{1}{2 [\sqrt{3} s p]}\sum_{l = p - [\sqrt{3} s p]}^{p + [\sqrt{3} s p]} \delta_{k,l} .
\end{align}
We take for example the calculation of the edge of the bulk of the eigenvalue spectrum for Fig. 4. The solid lines in any of the aforementioned figures can be calculated in a similar way. 

We begin with Eqs.~(13) of the main text, from which we obtain
\begin{align}
	\frac{1}{\sigma^2} &= \frac{1}{2 [\sqrt{3} s p]}\sum_{k = p - [\sqrt{3} s p]}^{p + [\sqrt{3} s p]}  \frac{(k/p)^2}{(\lambda_{\mathrm{edge}} - \Gamma \sigma^2 k A/p)^2 } , \nonumber \\
	A &=  \frac{1}{2 [\sqrt{3} s p]}\sum_{k = p - [\sqrt{3} s p]}^{p + [\sqrt{3} s p]}  \frac{k/p }{(\lambda_{\mathrm{edge}} - \Gamma \sigma^2 k A/p) }. 
\end{align}
For large $p$, these sums can be approximated by integrals such that we can write (using the substitution $x = k/p$)
\begin{align}
	\frac{1}{\sigma^2} &= \frac{1}{2 \sqrt{3} s }\int_{1 - \sqrt{3} s }^{1 + \sqrt{3} s } dx  \frac{x^2}{(\lambda_{\mathrm{edge}} - \Gamma \sigma^2 x A)^2 } , \nonumber \\
	A &=  \frac{1}{2 \sqrt{3} s }\int_{1 - \sqrt{3} s }^{1 + \sqrt{3} s } dx \frac{x }{(\lambda_{\mathrm{edge}} - \Gamma \sigma^2 x A) }. 
\end{align}
These are standard integrals that can be evaluated. One thus has to solve the following simultaneous equations numerically to find $A$ and $\lambda_{\mathrm{edge}}$ (letting $x_1 = 1 - \sqrt{3} s$ and $x_2 = 1 + \sqrt{3} s$ for shorthand)
\begin{align}
	\frac{1}{\sigma^2} =& \frac{1}{(x_2 - x_1) (\Gamma \sigma^2 A)^3} \Bigg\{\frac{\lambda_{\mathrm{edge}}^2}{\lambda_{\mathrm{edge}} - \Gamma \sigma^2 A x_2} - \frac{\lambda_{\mathrm{edge}}^2}{\lambda_{\mathrm{edge}} - \Gamma \sigma^2 A x_1} + \Gamma \sigma^2 A (x_2 - x_1)\nonumber \\
	& + 2 \lambda_{\mathrm{edge}} \ln\left[\frac{\lambda_{\mathrm{edge}} - \Gamma \sigma^2 A x_2}{\lambda_{\mathrm{edge}} - \Gamma \sigma^2 A x_1} \right] \Bigg\} , \nonumber \\
	A =&  -\frac{1}{(x_2 - x_1) (\Gamma \sigma^2 A)^2} \left\{ \lambda_{\mathrm{edge}} \ln\left[\frac{\lambda_{\mathrm{edge}} - \Gamma \sigma^2 A x_2}{\lambda_{\mathrm{edge}} - \Gamma \sigma^2 A x_1} \right] + A \Gamma \sigma^2 (x_2 - x_1) \right\} .
\end{align}
These simultaneous equations can be solved numerically to yield $\lambda_\mathrm{edge}$ as a function of (for example) $\Gamma$, as is plotted in Figs. 4 of the main text.

\end{document}